\documentclass[aps,prl,twocolumn,superscriptaddress,showpacs,floatfix]{revtex4-2}
\usepackage{amsmath,amssymb,amsfonts,float,graphics,epsfig,epstopdf,color,verbatim,tabularx,bm,multirow,appendix}
\usepackage[utf8]{inputenc}
\usepackage[T1]{fontenc}
\usepackage{xcolor}
\usepackage{dsfont}
\usepackage{textcomp}
\usepackage{yfonts}
\usepackage{footnote}
\usepackage{bm}
\usepackage{subfigure}
\usepackage{mathrsfs}
\usepackage{graphicx}
\usepackage{verbatim}
\usepackage{hyperref}
\usepackage{multirow}
\usepackage{braket}
\usepackage[normalem]{ulem}
\usepackage{tikz}
\usetikzlibrary{calc}
\usetikzlibrary{shapes.multipart}

\DeclareMathOperator{\Tr}{Tr}

\renewcommand{\vec}[1]{{\mathbf #1}}

\newcommand{\comments}[1]{}

\newcommand{\Sec}[1]{Sec.~\ref{#1}}

\newcommand{\Eq}[1]{Eq.~\eqref{#1}}

\newcommand{\Fig}[1]{Fig.~\ref{#1}}

\newcommand{\bcx}[1]{\textcolor{lightgray}{\ifmmode\text{\sout{\ensuremath{#1}}}\else\sout{#1}\fi}}

\newcommand{\stkout}[1]{\ifmmode\text{\sout{\ensuremath{#1}}}\else\sout{#1}\fi}

\newcommand\startsupplement{%
       \newpage\clearpage
       \setcounter{secnumdepth}{2}
       \setcounter{table}{0}
       \renewcommand{\thetable}{S\arabic{table}}
       \setcounter{figure}{0}
       \renewcommand{\thefigure}{S\arabic{figure}}
       \setcounter{equation}{0}
       \renewcommand{\theequation}{S\arabic{equation}}
       \setcounter{section}{0}
       \renewcommand{\thesection}{Section \Roman{section}}
       \renewcommand{\thesubsection}{\Roman{section}. \Alph{subsection}}
    }

\makeatletter
\def\l@subsubsection#1#2{}
\makeatother

\begin{document}

\title{Phases of (2+1)D SO(5) non-linear sigma model with a topological term on a sphere:\\ multicritical point and disorder phase}

\author{Bin-Bin Chen}
\affiliation{Department of Physics and HKU-UCAS Joint Institute of Theoretical and Computational Physics, The University of Hong Kong, Pokfulam Road, Hong Kong SAR, China}

\author{Xu Zhang}
\affiliation{Department of Physics and HKU-UCAS Joint Institute of Theoretical and Computational Physics, The University of Hong Kong, Pokfulam Road, Hong Kong SAR, China}

\author{Yuxuan Wang}
\email{yuxuan.wang@ufl.edu}
\affiliation{Department of Physics, University of Florida, Gainesville, FL 32601, USA}

\author{Kai Sun}
\email{sunkai@umich.edu}
\affiliation{Department of Physics, University of Michigan, Ann Arbor, MI 48109, USA}

\author{Zi Yang Meng}
\email{zymeng@hku.hk}
\affiliation{Department of Physics and HKU-UCAS Joint Institute of Theoretical and Computational Physics, The University of Hong Kong, Pokfulam Road, Hong Kong SAR, China}

\begin{abstract}
Novel critical phenomena beyond the Landau-Ginzburg-Wilson paradigm have been long sought after.
Among many candidate scenarios, 
the deconfined quantum critical point (DQCP) constitutes the most fascinating one, and its lattice model realization has been debated over the past two decades. 
Here we apply the spherical Landau level regularization 
upon the exact (2+1)D SO(5) non-linear sigma model with a topological term to study the potential DQCP therein. 
{
We perform density matrix renormalization group (DMRG) simulation with 
SU(2)$_\mathrm{spin}\times$U(1)$_\mathrm{charge}\times$U(1)$_\mathrm{angular-momentum}$ 
symmetries explicitly implemented. {Using } crossing point analysis for 
the critical properties of the DMRG data,
accompanied by quantum Monte Carlo simulations, {we} accurately obtain the comprehensive phase diagram of the model and find various novel quantum phases, 
including N\'eel, ferromagnet (FM), valence bond solid (VBS), valley polarized (VP) states 
and a gapless quantum disordered phase occupying extended area of the phase diagram. The VBS-Disorder and N\'eel-Disorder transitions are continuous with non-Wilson-Fisher exponents. 
Our results show the VBS and N\'eel states are separated by either a weakly first-order transition or the disordered region with a multicritical point in between, thus 
opening up more interesting questions on the two-decade long debate on the nature of DQCP.}
\end{abstract}

\date{\today}
\maketitle

\noindent{\textcolor{blue}{\it Introduction.}--}
Over the past two decades, the enigma of the deconfined quantum critical point (DQCP) has never failed to attract attention across the communities of condensed matter to quantum field theory and high-energy physics, as it is believed to offer a new paradigm in theory~\cite{senthilQuantum2004,nahumDeconfined2015,qinDuality2017,wangDeconfined2017,senthilDeconfined2023}, numerical simulation~\cite{sandvikEvidence2007,louvbsneel2009,haradaPossibility2013,liuSuperconductivity2019,liaoDiracI2022,shaoQuantum2016,maDynamics2018,nahumEmergent2015,sreejithEmergent2019,maRole2019}, and experiment~\cite{jimenezquantum2021,zayed4spin2017,guoQuantum2020,sunEmergent2021,cuiProximate2023,guoDeconfined2023} that goes beyond the Landau-Ginzburg-Wilson (LGW) framework of phase transitions. 

However, the lattice realizations of DQCP have been debated ever since. In SU(2) spin systems, the $J$-$Q$ model~\cite{sandvikEvidence2007} was initially believed to realize a DQCP between N\'eel and valence bond solid (VBS) states. Over the years, a plethroa of results have been reported, including the emergent continuous symmetry with fractionalized excitations~\cite{nahumEmergent2015,maRole2019,sreejithEmergent2019,maDynamics2018} yet drifting critical exponents incompatible with conformal bootstrap bounds (with one O(3)$\times\mathbb{Z}_4$ singlet)~\cite{haradaPossibility2013,nahumDeconfined2015,shaoQuantum2016,nakayamaNecessary2016,polandConformal2019}, weakly first-order pseudocriticality versus continuous transition or multicritical point~\cite{kuklovDeconfined2008,jiangFrom2008,chenDeconfined2013,zhaoMulticritical2020,demidioDiagnosing2021,maTheory2020,nahumNote2020,nahumDeconfined2015,sandvikConsistent2020}, and violation of entanglement positivity for a unitary conformal field theory (CFT)~\cite{zhaoScaling2022,wangScaling2022,liuDisorder2023}, 
and debate regarding the nature of the phase transition persists to this day. 
A more recent quantum Monte Carlo (QMC) study suggests the non-unitary CFT of the DQCP scenario in SU($N$) spin systems for $N<N_c\simeq8$~\cite{song2023deconfined}.

Similar changing perceptions also occur in DQCP models with fermions, realizing transitions from a Dirac semimetal (DSM)  through quantum spin Hall insulator to superconductor~\cite{liuSuperconductivity2019,liuFermion2023,liuMetallic2022}, or from DSM through VBS to N\'eel state~\cite{liaoDiracI2022,liaoDiracIII2022,liaoTeaching2023}. The inclusion of fermions offers advantages over the previous model, due to the absence of symmetry-allowed quadruple monopoles and the associated
second length scale that breaks the assumed U(1) symmetry down to $\mathbb{Z}_4$~\cite{shaoQuantum2016,nahumDeconfined2015}, but the non-compatible critical exponents still persist and the accumulating numerical results are also pointing towards a non-unitary CFT of these DQCPs~\cite{liuSuperconductivity2019,wangDoping2021,wangPhases2021,liaoDiracI2022,liuFermion2023,liaoTeaching2023,liuDisorder2023}. Despite extensive efforts over
the past two decades, the lattice realizations of DQCP in its original sense of beyond LGW and yet still critical, with emergent continuous symmetry and fractionalized excitations, are still in ``The Enigma of Arrival''~\cite{naipaulEnigma2020}.

A key origin of the debate stems from the fundamental requirement of emergent symmetries at DQCPs. For instance, the $J$-$Q$ model DQCP { is speculated to have a U(1) symmetry emerge out of the $\mathbb{Z}_4$ symmetry of VBS, which is then speculated to be combined with the SU(2) symmetry of the N\'eel order to give rise to the ultimate SO(5) emergent symmetry.} Due to the extremely slow RG flow towards such emergent symmetries, numerical studies face
challenges in accessing these {speculated} DQCPs due to finite size effects. To overcome this challenge, lattice models with explicit SO(5) symmetry have been introduced, e.g., the (2+1)D SO(5) nonlinear sigma model (NLSM) with a Wess-Zumino-Witten (WZW) topological term~\cite{leeWess2015}. {In such a model, different from the aforementioned $J$-$Q$ and fermion realizations, one can directly ask the question whether there is a continuous N\'eel-VBS transition in its phase diagram, without the hierarchy of symmetries emergence.} 

However, previous attempts {for such a SO(5) model} with the half-filled Landau level of Dirac fermions as a regularization on torus geometry, are unfortunately limited by severe computational complexity both for density matrix renormalization group (DMRG) and QMC simulations~\cite{impoliteHalf2018,wangPhases2021}. {Moreover, these works have not addressed the entire phase digram with control parameters moving away from the SO(5) symmetric path, such that the transitions towards the SO(3) symmetry-breaking N\'eel phase and the SO(2) symmetry-breaking VBS phase have not been addressed. Therefore, the results are still inconclusive and different scenarios---such as the first order transition between N\'eel and VBS phases, the multi-critical point and the DQCP scenarios---are all suggested.} 

\begin{figure}[t!]
	\includegraphics[width=\columnwidth]{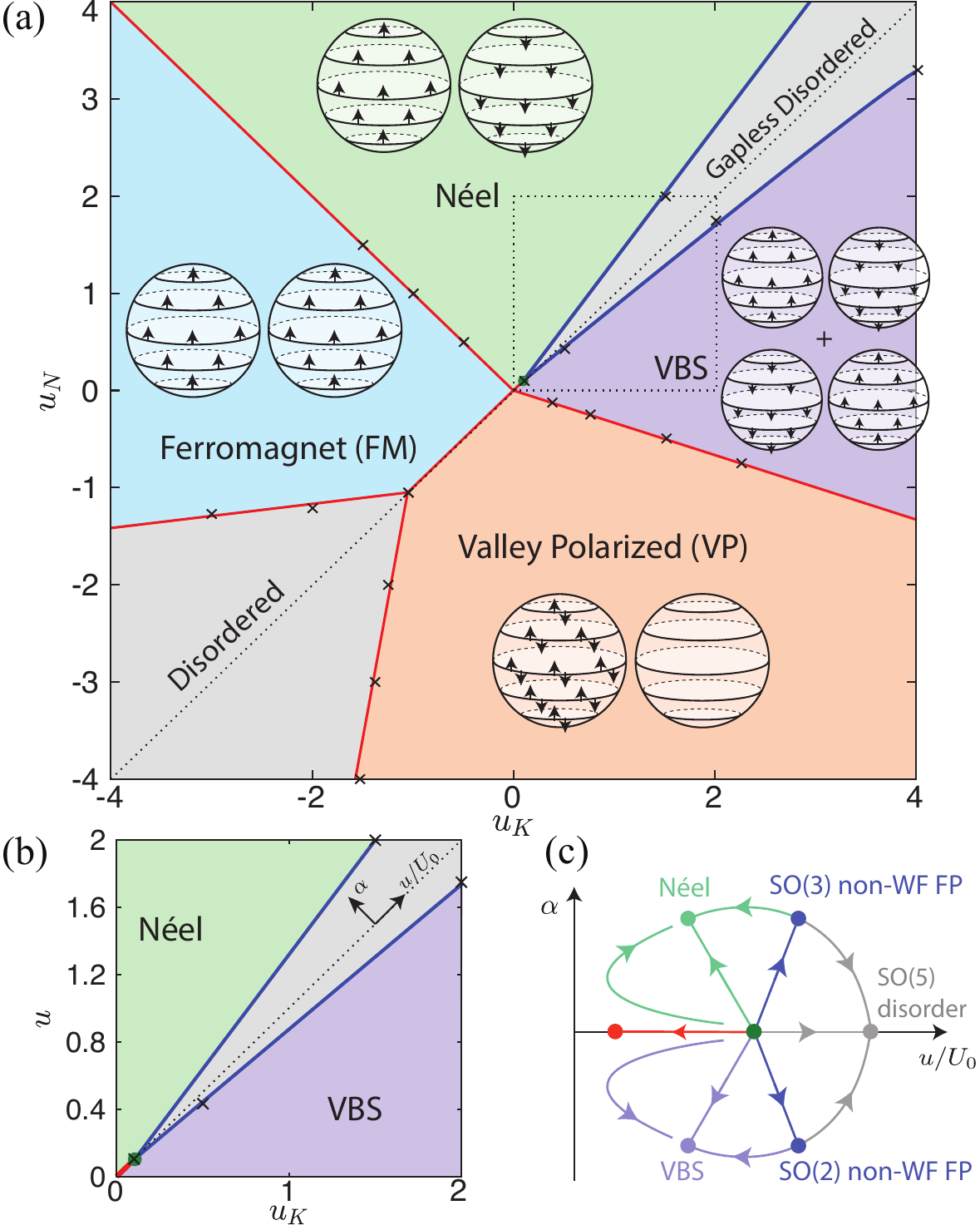}
	\caption{\textbf{The groundstate phase diagram and RG flow of the SO(5) model.} 
		(a) Overall phase diagram with
		N\'eel, VBS, ferromagnet (FM), valley 
		polarized (VP) phases, and the disorder phases as denoted.
		The deep blue lines denote the { continuous and the non-Wilson-Fisher} transition, 
		the red lines denote the first-order transition, and the deep green 
		dot denotes a multicritical point. 
		The 4 symmetry-breaking states are schematically depicted by two spheres for the two opposite valleys, and the 
		spin degrees of freedom are depicted by the arrow directions.
		(b) Zoomed-in phase diagram as indicated by the dashed box in panel (a). 
		The two critical boundaries meet at a multicritical point (deep green dot) 
		below which the SO(5) symmetry is spontaneoulsy broken.
		(c) Possible RG flow in the considered parameter space in (b), with 
		multicritical point (deep green dot), SO(5) disorder (grey dot), { non-Wilson-Fisher fixed points (blue dots) towards SO(2) breaking VBS-ordered (light purple dot) and SO(3) breaking N\'eel-orderd (light green dot) fixed points, and the SO(5) breaking (red dot) fixed point.}
		The $\alpha$ and $u/U_0$ axes are indicated in panel (b).
	}
	\label{fig:fig1}
\end{figure}

Here, we push forward the solution of the problem by applying the spherical Landau level regularization which was studied in the context of fractional quantum Hall effect in early literatures~\cite{haldaneFractional1983,sondhiSkyrmions1993,melikQuantum1997,zhaoFractional2011} and has recently been shown to suffer less finite-size effect than the torus geometry for the (2+1)D Ising model~\cite{zhuUncovering2023}. {To facilitate the large system sizes and quantitative data analysis, we perform DMRG simulation with 
explicit SU(2)$_\mathrm{spin}\times$U(1)$_\mathrm{charge}\times$U(1)$_\mathrm{angular-momentum}$ symmetries symmetries, accompanied with exact diagonalization (ED) and QMC simulations. 
We accurately simulate the entire phase diagram of the model with various novel quantum states identified, including the N\'eel, VBS, ferromagnet (FM), and valley polarized (VP) states. Most importantly, we find a {\it gapless disordered region} separates the VBS and N\'eel states. We employ the crossing point analysis~\cite{luckCorrections1985,qinDuality2017,shaoQuantum2016,maAnomalous2018,maRole2019} for the critical properties of the DMRG data and find the VBS-Disorder and N\'eel-Disorder transitions are continuous with non-Wilson-Fisher exponents. {\color{black} These critical boundaries meet at a multicritical point along the SO(5) line behind which the SO(5) symmetry is explicitly broken with weakly first order transition between the N\'eel and VBS phases. Our results are supported by recent conformal boostrap analysis on the quantum tricriticality on the DQCP~\cite{chesterBootstrapping2023}, as well as the QMC entanglement entropy results of the JQ model that at the N\'eel-VBS transition is weakly first order~\cite{songExtracting2023,dengDiagnosing2024}.}

{Our discovery of the extended gapless disordered phase and the multicritical point 
and  our novel methodology 
of the crossing point analysis of the DMRG data,
open a few new research directions, such as the nature of the disordered phase, its relation with pseudo-criticality
and symmetry-enforced gaplessness~\cite{wangDeconfined2017}, and its transition between VBS and N\'eel phases. {These results} substantially {advance} the two-decade long quest of DQCP in the phase diagram of the (2+1)D SO(5) NLSM with WZW topological term.}

\noindent{\textcolor{blue}{\it Model and Methods.}--}
We consider the (2+1)D Hamiltonian
$H_\Gamma = \frac{1}{2}\int d\Omega \{U_0 \left[\psi^\dag(\Omega)\psi(\Omega)-2\right]^2 -\sum_{i=1}^5 u_i \left[\psi^\dag(\Omega)\Gamma^i\psi(\Omega)\right]^2\}$, 
where $\psi_{\tau\sigma}(\Omega)$ is the 4-component Dirac fermion annihilation operator 
with valley $\tau$ and spin $\sigma$ indices, and 
$\Gamma^i=\{\tau_x\otimes\mathbb{I}, \tau_y\otimes\mathbb{I}, \tau_z\otimes{\sigma_x}, 
\tau_z\otimes{\sigma_y}, \tau_z\otimes{\sigma_z}\}$ 
are the 5 mutually anticommuting matrices, whose commutators 
$L^{ij}=\tfrac{-i}{2}[\Gamma^i,\Gamma^j]$ are generators of the SO(5) group. 
Subsequently, we project the Hamiltonian onto the zero energy Landau level on the sphere, which is the same as the lowest massive fermion Landau levels (LLL) of a sphere with $4\pi s$ magnetic monopole~\cite{JELLAL2008,Arciniaga2016,GREITER2018}, where the $(2s+1)$-fold degenerate 
LLL wavefunctions are 
$\Phi_m(\Omega)\propto e^{im\phi}\cos^{s+m}(\tfrac{\theta}{2})\sin^{s-m}(\tfrac{\theta}{2})$ 
with $m\in\{-s, -s+1,\cdots,s\}$ and $2s \in\mathbb{Z}$. 
Via the expansion 
$\psi(\Omega)=\sum_m \Phi_m(\Omega)c_m$, we have
\begin{align}\label{Eq:Model}
\begin{aligned}
&\hat H_\Gamma = U_0 \hat H_0 - \sum_i u_i \hat H_i, \text{with} \\
&\hat H_i=\sum_{m_1,m_2,m}V_{m_1,m_2,m_2-m,m_1+m}\times\\
& \left(c^\dag_{m_1}\Gamma^i c^{\,}_{m_1+m}-2\delta_{i0}\delta_{m0}\right)
 \left(c^\dag_{m_2}\Gamma^i c^{\,}_{m_2-m}-2\delta_{i0}\delta_{m0}\right)
 \end{aligned}
\end{align}
with $\Gamma^0=\mathbb{I}\otimes\mathbb{I}$. 
The precise form of $V_{m_1,m_2,m_3,m_4}$ 
can be found in Supplementary Materials (SM)~\cite{suppl}.
Throughout, we set $U_0=1$ as the energy unit and 
let $u_1=u_2=u_K, u_3=u_4=u_5 = u_N$. 
{When $u_K=u_N>0$, this model is known to be described by a SO(5) NLSM with WZW term}~\cite{leeWess2015,impoliteHalf2018,wangPhases2021}.
When $u_K \neq u_N$, the symmetry reduces to SO(3)$\times$SO(2).  
For positive $u_{K,N}$, it was proposed that $u_N > u_K$ stabilizes the N\'eel
order, which spontaneously breaks the SO(3) symmetry,
while $u_N < u_K$ {favors a valley order} breaking the SO(2)
symmetry, which in a lattice model can be interpreted as the VBS order. {We note, however, such explicit perturbation away from the SO(5) symmetric path have not been investigated in previous studies.}
If a direct {and continuous} phase transition between these two
states arises at $u_K = u_N$, at the transition the system has an
explicit SO(5) symmetry, which realizes a DQCP. 
While {previous works mainly focused on positive values for $u_{K,N}$ along the SO(5) line, we sweep the entire $(u_K,u_N)$ plane for symmetry breaking phases.}

We perform DMRG simulation with SU(2)$_\mathrm{spin}\times$U(1)$_\mathrm{charge}${$\times$U(1)$_\mathrm{angular-momentum}$} symmetries in the tensor library QSpace~\cite{Weichselbaum2012,Weichselbaum2020,Bruognolo2021}, and 
keep up to {$4096$} SU(2) invariant multiplets (equivalent to $\sim{12000}$ 
U(1) states) to render the truncation errors within $5\times10^{-5}$. 
We also perform determinant QMC as well as ED
simulations as complements. 
We denote the system size by the Landau level degeneracy $N=2s+1$ and obtain converging 
results up to $N={16}$, the largest size achieved so far for the model on sphere to our knowledge. { To determine the VBS-disorder and N\'eel-disorder critical points and the critical exponents in an unbiased manner, we adopt the crossing point analysis {that has been used in earlier studies for many in quantum-critical spin models~\cite{luckCorrections1985,qinDuality2017,shaoQuantum2016,maAnomalous2018,maRole2019}. The derivation and detailed steps are given in SM~\cite{suppl}.}}

\noindent{\textcolor{blue}{\it Phase Diagram.}--}
We first give a summary of the phase diagram. For
all the ordered phases observed, the order parameters take the form of fermion bilinears: $\langle O \rangle =
\int d\Omega \langle \psi^{\dagger}(\Omega)M \psi(\Omega)\rangle = \sum_{m}
\langle c^{\dagger}_m M c_m\rangle $, where $M$ is either a $\Gamma$-matrix or one of the SO(5)
generators $L^{ij}$.
In the case of $(u_K, u_N)>0$,  
there are 3 phases including the N\'eel state 
(ordered in the $\Gamma^{3,4,5}$ directions), 
the VBS (ordered in the $\Gamma^{1,2}$ directions), 
and the disorder phase, as shown in \Fig{fig:fig1}.  
At small $u_{K,N}$ (below $\sim0.1$), the N\'eel and
VBS phases are separated by a first-order phase boundary,
along the $u_K = u_N$ line with SO(5) symmetry.
At large $u_{K,N}$, instead of the proposed
direct {and continuious} transition, we find that N\'eel and VBS phases are
separated by an intermediate disordered phase, and continuous
transitions from the disordered state to both N\'eel and VBS states {(We will discuss the {critical behavior} of VBS-disorder and N\'eel-disorder transitions in the next section.)}

For {negative values of $u_K$ and/or $u_N$,} we {find} 3 phases: the FM state ($M = L^{34},L^{35},L^{45}$)
where both valleys exhibit the same magnetization direction,
the VP state ($M = L^{12}$) which breaks an Ising $\mathbb{Z}_2$
symmetry, and {another} disorder phase. 
When $|u_{K,N}|$ are small (i.e., $(u_K, u_N) > -1$),
the FM and VP states are directly connected by a first-order
transition along the SO(5) line. {Again, for larger $|u_{K,N}|$}, 
the FM and VP phases are separated
by the disordered phase, {while} the transitions between
the FM/VP states and the disordered state are all first-order.

The transition between the FM and N\'eel states takes
place in the quadrant of $u_K < 0$ and $u_N > 0$ through a
first order phase boundary. Similarly, a first-order transition
between the VBS and VP states is observed in the quadrant of
$u_K > 0$ and $u_N < 0$.

\begin{figure}[t!]
\includegraphics[width=\columnwidth]{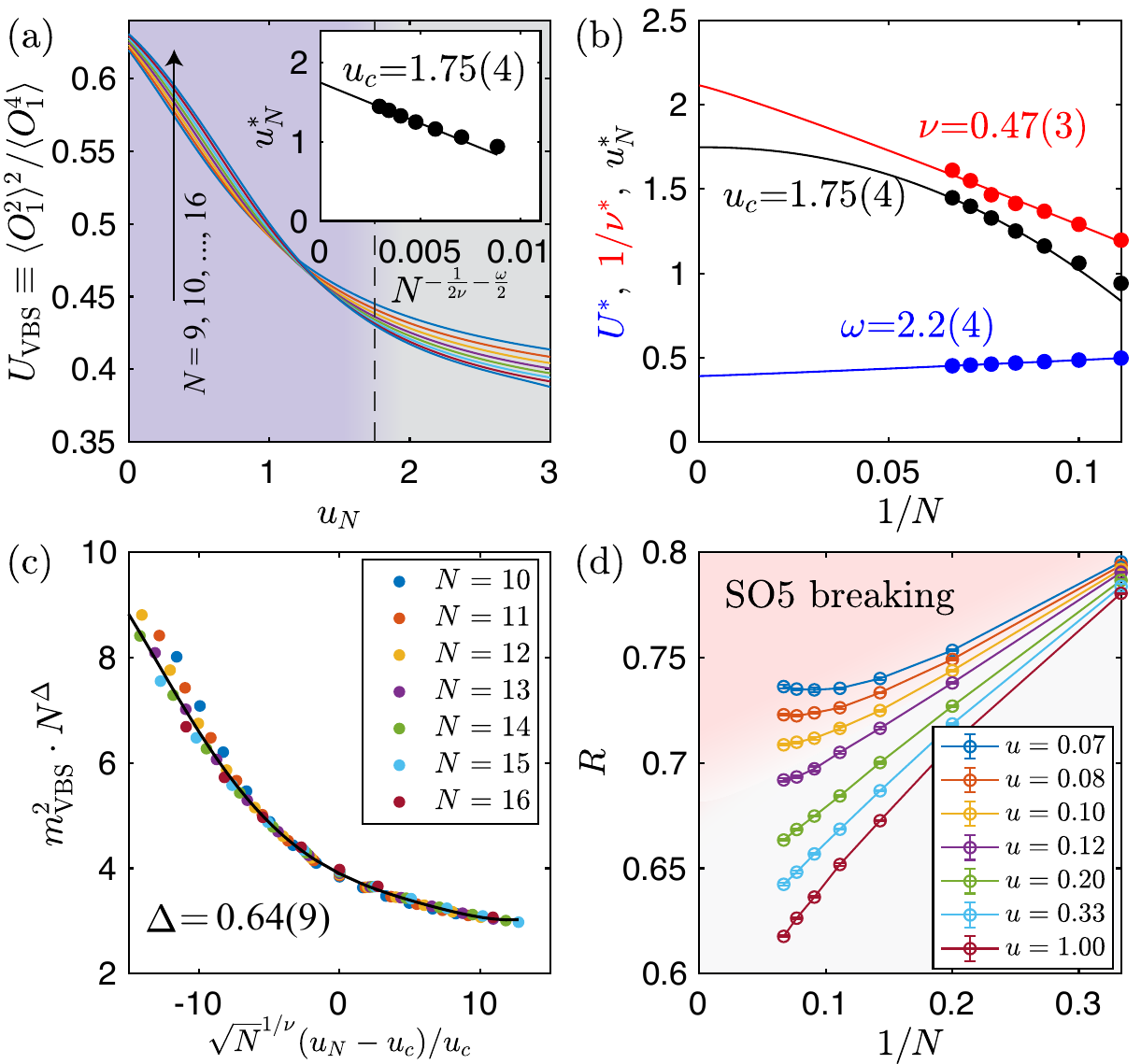}
\caption{\textbf{Crossing point analysis of the VBS-disorder transitions.} 
Along the fixed $u_K=2$ cut, 
(a) the VBS Binder ratio $U_\mathrm{VBS}\equiv\langle O^2_1\rangle^2/\langle O^4_1\rangle$ crosses between
 successive size pair $(N,N+1)$, whose crossing points $u_N^\ast$ drift towards larger $u_N$ 
 with larger $N$. In the inset, $u_N^\ast$'s are extrapolated to $u_c=1.75(4)$ in the thermodynamic 
 limit with the scaling form $u_N^\ast(N) = u_c + N^{-\frac{1}{2\nu}-\frac{\omega}{2}}$, 
 {with $\nu=0.47(3)$ and $\omega=2.2(4)$ from the crossing point analysis shown in SM~\cite{suppl}.} 
 (b) The subleading operator exponent $\omega$, correlation length exponent $\nu$, and 
 the critical point $u_c$ is obtained from 
 the scaling form of crossing point, Binder ratios value at crossing point 
 and its first-order derivatives.
(c) $m^2_\mathrm{VBS}$ rescaled by $N^{\Delta}$ with scaling dimension $\Delta=\color{black}\color{black}0.64(9)$ 
versus $\sqrt{N}^{1/\nu} (u_N-u_c)/u_c$, collapses nicely for various system sizes $N=9,10,...,16$.
(d) Correlation ratio $R$ (up to $N=15$), 
along the SO(5) line, indicate the phase transition point near $u\simeq0.1$.}
\label{fig:fig2}
\end{figure}

\noindent{\textcolor{blue}{\it Phases of $(u_K, u_N)>0$ quadrant.}--}
We first focus on the positive $u_{K,N}$ cases, 
and compute the squared order parameter $\langle O_i^2\rangle$ with 
$O_i = \int d\Omega \psi^\dag(\Omega)\Gamma^i\psi(\Omega) = \sum_m c^\dag_m \Gamma^i c^{\,}_m$.
We use $m^2_\text{N\'eel} = \tfrac{1}{3N^2}\langle(O_3^2+O_4^2+O_5^2)\rangle$ and $m^2_\mathrm{VBS} = \tfrac{1}{2N^2}\langle(O_1^2+O_2^2)\rangle$
for N\'eel and VBS orders, respectively. 

{To systematically determine the VBS-Disorder transition, we fix a few $u_K=0.5,2,4$ values and scan $u_N$. The representative $u_K=2$ scan are shown in Fig.~\ref{fig:fig2}. Fig.~\ref{fig:fig2} (a) show the VBS Binder ratio $U_\mathrm{VBS}\equiv\langle O^2_1\rangle^2/\langle O^4_1\rangle$ crosses between successive size pair $(N,N+1)$, it is clear that there is a crossing of the data which indicates the transition point. To locate the transition point in an unbiased manner, we employ the crossing point analysis as detailed in SM~\cite{suppl} and find that the $u^{*}_N = u_{c}+N^{-\frac{1}{2\nu}-\frac{\omega}{2}}$ 
(the asterisk indicates the finite-size 
crossing points)
nicely extrapolate to the $u_c=1.75(4)$ with the correlation length exponent 
$\nu=0.47(3)$ and subleading exponent $\omega=2.2(4)$ independently obtained from Binder ratio 
$U^\ast(u_N^\ast, N) = a + b N^{-\frac{\omega}{2}}$
and its derivatives
$\frac{1}{\nu^\ast}\equiv{2N}\ln{\frac{U'(u_N^\ast,N+1)}{U'(u_N^\ast,N)}} = \frac{1}{\nu} - c N^{-\frac{\omega}{2}}$
at finite $N$, as shown in Fig.~\ref{fig:fig2} (b). 
With the obtained $u_c$ and $\nu$, one can further collapse the VBS order parameter 
as $m^{2}_\mathrm{VBS}\cdot N^{\Delta_\mathrm{VBS}}$ against $\sqrt{N}^{1/\nu}(u_N-u_c)/u_c$ 
and unbiasedly obtain the scaling dimension $\Delta_\mathrm{VBS}=\color{black}0.64(9)$, 
as shown in Fig.~\ref{fig:fig2} (c). 
We note the collapse is of very good quality and the obtained 
$\Delta_\mathrm{VBS}=\color{black}0.64(9)$ is substantially larger than its O(2) 
Wilson-Fisher counterpart 0.519. 
This gives a clear signature, that the VBS-Disorder transition is not of 
Wilson-Fisher type and there is no direct VBS-N\'eel DQCP transition at $u_K=2$. 
We have further performed the same analysis at $u_K=0.5,4$ and obtained equally good 
and consistent critical point $u_c=0.43(3), 3.3(2)$ and exponents 
$\nu=0.55(5), 0.49(5)$, $\omega=2.2(4), 2.1(1)$ and 
VBS scaling dimension $\Delta_\text{VBS}=\color{black}0.63(8), \color{black}0.63(9)$, 
the results are shown in SM~\cite{suppl}.} 
{Similar simulations are performed with fixed $u_N=2$ cut and 
$u_c=1.5(3), \Delta=\color{black}0.55(3)$ are found. With these data, 
we map out the phase boundaries of both VBS-Disorder and 
N\'eel-Disorder transitions as shown in \Fig{fig:fig1} (a).
We find the continuous VBS-Disorder and N\'eel-Disorder transitions are merged into one multicritical point at $u_K=u_N\simeq0.1$, as denoted in Fig.~\ref{fig:fig1} (b).
For $u_K=u_N=u\lesssim0.1$, the SO(5) line {represents a first-order phase boundary} with 
SO(5) symmetry spontaneously broken.}

To verify such a first-order line, we simulate along the exact SO(5) line 
$u_K=u_N=u$. As shown in \Fig{fig:fig2} (d), 
correlation ratios
$R\equiv1-\left\langle \mathbf{O}^2_{l=1}\right\rangle/\left\langle \mathbf{O}^2_{l=0}\right\rangle$ 
for up to sizes $N=15$,
indicate the phase transition point near $u\simeq0.1$. 
Here $\mathbf{O}_{l}\equiv(O_{1,l}, \cdots,O_{5,l})$ is the O(5) order parameter with angular momentum shift $l$~\cite{suppl}.
Since the multicritical point is the meeting point of the SO(2)-breaking and SO(3)-breaking critical boundaries, it requires to fine-tune two different control parameters in order to access. 

{
Within disordered phase, we calculate the spin-singlet gap 
$\Delta_0 = E_1(S=0) - E_0(S=0)$ and triplet gap 
$\Delta_1 = E_0(S=1) - E_0(S=0)$, with $E_i(S)$ the $i$-th lowest energy 
in the total spin-$S$ sector.
In \Fig{fig:fig3}, both kinds of gaps follow a clear $1/\sqrt{N}$ behaviour and 
scale to zero in the thermodynamic limit. 
Such scaling behavior of gaps strongly imply the disordered phase is gapless, fully consistent with symmetry-enforced gaplessness discussed in Ref.~\cite{wangDeconfined2017}.}

\begin{figure}[t!]
\includegraphics[width=\columnwidth]{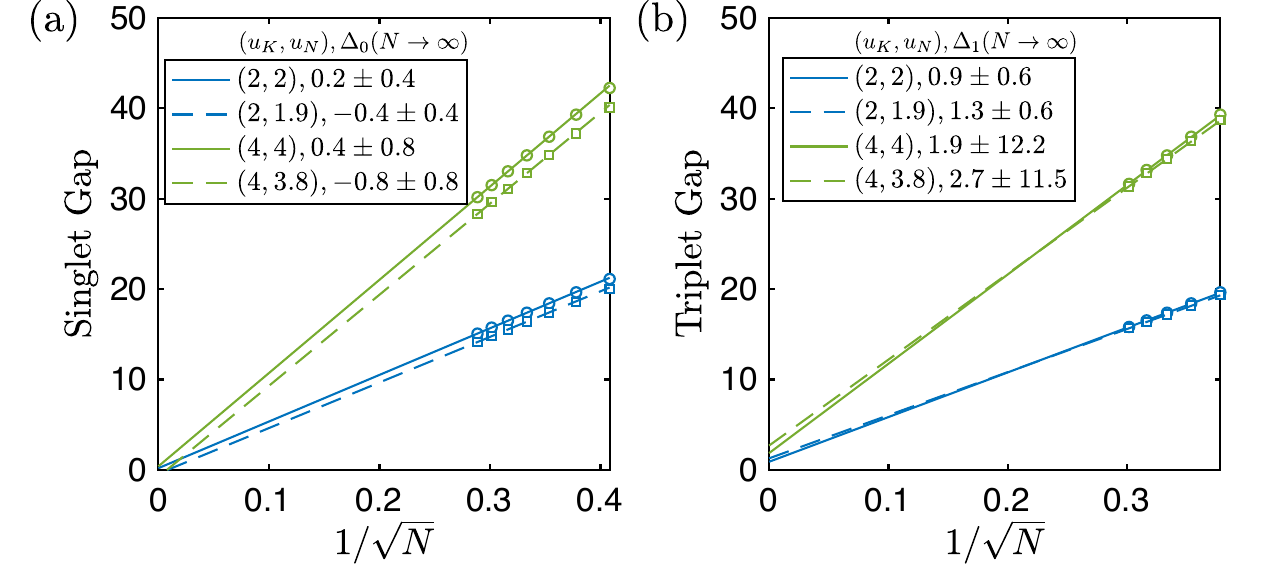}
\caption{\textbf{Spin gaps within disordered phase.} 
Within the disordered phase, (a) spin singlet gaps, (b) spin triplet gaps are calculated 
in the finite-size cases, and extrapolated to zero with $1/\sqrt{N}$ in the thermodynamic limit. 
}
\label{fig:fig3}
\end{figure}

\begin{figure}[t!]
\includegraphics[width=\columnwidth]{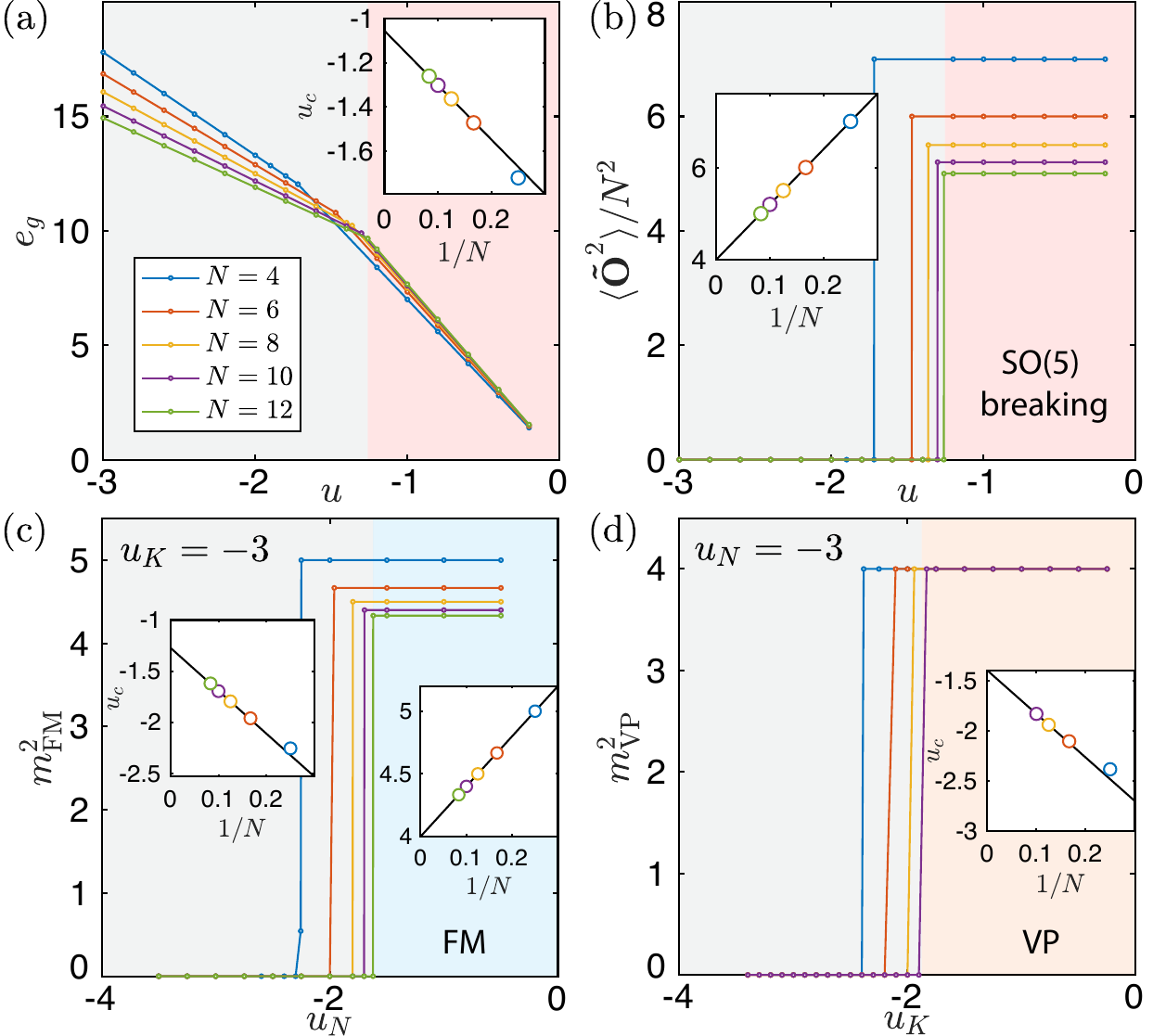}
\caption{\textbf{Identification of the FM and VP phases.} 
Along the negative SO(5) line $u_N=u_K=u$, 
(a) the groundstate energy $e_g$ exhibits kinks around $u\sim-1$. 
In the inset, the kink positions $u_c(N)$ is extrapolated to $u_c(\infty)\simeq-1.056$. 
(b) $\langle {\bf\tilde O}^2\rangle/N^2$ shows a sudden jump around $u_c$.
(c) Along a fixed $u_K=-3$ line, $m^2_\mathrm{FM}$ shows a sudden jump behaviour. 
In the left inset, the transition points are extrapolated linearly to $u_c(\infty)\simeq-1.25$. 
(d) Along a fixed $u_N=-3$ line, $m^2_\mathrm{VP}$ shows a
sudden jump behaviour.
In the inset, the transition points are extrapolated linearly to
$u_c(\infty)\simeq-1.38$. 
}
\label{fig:fig4}
\end{figure}

\noindent{\textcolor{blue}{\it Phases of $(u_K, u_N) <0$ quadrant.}--}
For negative $u_K$ and $u_N$, the order parameter with $M=\Gamma^{i}$ vanishes in the thermodynamic limit. 
Instead, the relevant order parameter involves the SO(5) generator $M=L^{ij}$. We calculate the squared generators $\langle \tilde O_{ij}^2\rangle$ with
$ \tilde O_{ij} = \int d\Omega \psi^\dag(\Omega)L^{ij}\psi(\Omega) = \sum_m c^\dag_m L^{ij} c^{\,}_m$, 
and define the squared FM order parameter as 
$m^2_\mathrm{FM}=\tfrac{1}{N^2}\langle (\tilde O_{34}^2 + \tilde O_{35}^2 + \tilde O_{45}^2)\rangle$, 
and the squared VP order parameter as 
$m^2_\mathrm{VP}=\tfrac{1}{N^2}\langle \tilde O_{12}^2 \rangle$. 
As $L^{12}=\tau_z, L^{34}=-\sigma_z,  L^{35}=\sigma_y,  L^{45}=\sigma_x $, 
the finite value of $m^2_\mathrm{VP}$  and $m^2_\mathrm{FM}$
suggests the VP and FM states respectively.

In \Fig{fig:fig4}(a) and (b), we simulate along the negative SO(5) line $u_K=u_N=u<0$. 
The groundstate energies 
$e_g = \tfrac{1}{N}\langle\psi| H_\Gamma|\psi\rangle$ 
show clear kinks at $u_c(N)$ which can be extrapolated to $u_c(\infty)\simeq -1.056$ 
(c.f. the inset).
As shown in \Fig{fig:fig4}(b), 
such a first-order transition can also be seen from the squared order parameter
$\langle {\bf\tilde O}^2\rangle/N^2$,
which rapidly jumps from zero to a finite plateau,
whose heights decreases upon increasing $N$ and can be extrapolated to the value of $4$.
In \Fig{fig:fig4}(c), we further determine the $u_K=-3$ cut in the phase diagram. 
The value of $m^2_\mathrm{FM}$ jump from zero to finite around 
$u_N(N=\infty)\simeq-1.25$.
Similarly in \Fig{fig:fig4}(d), we simulate on fixed $u_N=-3$ cut where $m^2_\mathrm{VP}$ 
jump to finite value of $4$ around $u_K(N=\infty)\simeq-1.38$. 
{More DMRG results concerning the first-order transitions between VP and VBS, and between FM and N\'eel phases, are shown in SM~\cite{suppl}.}

{\noindent{\textcolor{blue}{\it Discussions.}}--}
Our study provides a comprehensive
phase diagram for the (2+1)D SO(5) NLSM with WZW term on a sphere. It
reveals novel quantum states and suggests a SO(5) disordered region separating the SO(2)
breaking VBS and SO(3) breaking N\'eel phases, which terminates at a multicritical
point~\cite{zhaoMulticritical2020}. 
{Our 
discovery of the extended disordered phase and the multicritical point {using a novel method } of  crossing point analysis of the DMRG data, may also offer a platform for the search of the predicted pseudo-critical behavior~\cite{wangDeconfined2017}, which we leave for future studies.}
These results, combined with recent observations
of {\color{black} weakly first-order transition from entanglement measurements~\cite{zhaoScaling2022,wangScaling2022,liuFermion2023,liaoTeaching2023,song2023deconfined,liuDisorder2023,songExtracting2023,dengDiagnosing2024} as well as the conformal boostrap deconfined quantum tricriticality~\cite{chesterBootstrapping2023}},  open up new directions for the two-decade long pursuit of DQCP 
in various N\'eel-to-VBS settings. 

Furthermore, our results find resonance with the experiments both in the VBS-AFM transition in quantum magnet SrCu$_2$(BO$_3$)$_2$~\cite{zayed4spin2017,guoQuantum2020, jimenezquantum2021,cuiProximate2023,guoDeconfined2023} and the QSH-SC transition in monolayer WTe$_2$~\cite{songUnconventional2023}, where the systems either exhibit a first order transition or an intermediate phase. A new pathway
towards conformal 2D SU(2) DQCP is recently proposed, with SO(5)$_f \times$ SO(5)$_b$ global symmetry~\cite{christosModel2023}. 
Investigating the validity of this newly proposed DQCP using present techniques would be of great interest. 

\begin{acknowledgments}
{\it{Note Added:-}} Recently, Ref.~\cite{zhouThe2023} reported pseudo-critical behavior for the SO(5) line. The parameter range of the reported pseudo-critical behavior and (approximate) conformal symmetry, i.e. $0.7<V/U<1.5$, correspond to {\color{black} $0.1187<u/U_0<0.4286$} 
{close to the multicritical point} in our phase diagram.

{\it{Acknowledgment.-}} We thank Subir Sachdev, Fakher Assaad, Meng Cheng, Yin-Chen He, Wei Zhu and Cenke Xu for valuable discussions on the related topic. BBC, XZ and ZYM thank Wei Zhu for fruitful discussion on spherical Landau level regularization. They acknowledge the Research Grants Council
(RGC) of Hong Kong Special Administrative Region of
China (Project Nos. 17301721, AoE/P701/20, 17309822,
C7037-22GF, 17302223), the ANR/RGC Joint Research
Scheme sponsored by RGC of Hong Kong and French
National Research Agency (Project No. A\_HKU703/22). YW is supported by NSF under award number DMR-2045781. The authors thank the Beijng PARATERA Tech CO.,Ltd.
(URL: https://cloud.paratera.com), the HPC2021 system under the Information Technology Services
for providing HPC resources that have contributed to the research results reported within this paper.
\end{acknowledgments}

\bibliography{bibtex}
\bibliographystyle{apsrev4-2}

\startsupplement

\begin{widetext}
\begin{center}
{\bf \uppercase{Supplemental Materials for \\[0.5em]
Phases of (2+1)D SO(5) non-linear sigma model with a topological term on a sphere:\\ multicritical point and disorder phase}}
\end{center}

\vskip3em

In Supplementary Materials \ref{sec:I}, we explain the spherical Landau level regularization of the SO(5) model. In \ref{sec:II}, we show the DMRG implementation of the model {with SU(2)$_\mathrm{spin}\times$U(1)$_\mathrm{charge}\times$U(1)$_\mathrm{angular-momentum}$ 
symmetries. In \ref{sec:III}, we show the ED and QMC implementation of the model as well as 
benchmark results of ED and QMC. 
In \ref{sec:iv}, we derive the finite size scaling equations for the crossing point analysis of the DMRG data. We note this is the first time such unbiased analysis of the critical properties (the critical point and exponents) has been introduced to the DMRG literature. 
In \ref{sec:v}, we show the lowest-lying states calculation of different fixed-particle-number sectors 
within the  phase separation region in the third quadrant. 
In \ref{sec:vi}, we show more DMRG results for the phase transitions in the phase diagram of  Fig.1 in the main text.}

\section{{Spherical Landau level regularization of SO(5) model}}
\label{sec:I}
\subsection{More on the SO(5) model}
Our notation is based on that used in Refs.~\cite{impoliteHalf2018,wangPhases2021,zhuUncovering2023}. 

We would like to project the SO(5) Hamiltonian onto the lowest Landau level (LLL) of the Haldane sphere. 
The original Hamiltonian is
\begin{equation}
H_{\Gamma} = \frac{1}{2}{\int{d\Omega_{1}{\int{d\Omega_{2}\delta\left( \left| {\Omega_{1} - \Omega_{2}} \right| \right){\sum\limits_{i = 0}^{5}{U_{i}\left( {\psi^{\dagger}\left( \Omega_{1} \right)\Gamma^{i}\psi\left( \Omega_{1} \right) - C\left( \Omega_{1} \right)\delta_{i,0}} \right)\left( {\psi^{\dagger}\left( \Omega_{2} \right)\Gamma^{i}\psi\left( \Omega_{2} \right) - C\left( \Omega_{2} \right)\delta_{i,0}} \right)}}}}}}
\end{equation}
where $\psi_\alpha(\Omega)$ is 4-component fermion annihilation operator with mixing valley and spin index $\alpha$, 
and $\Gamma^i=\{\tau_x\otimes\mathbb{I}, \tau_y\otimes\mathbb{I}, \tau_z\otimes\vec\sigma_x, \tau_z\otimes\vec\sigma_y, \tau_z\otimes\vec\sigma_z\}$ are the 5 mutually anticommuting matrices. 
Here, $U_i=-u_i$ for $i\neq0$ as shown in the main text and this will be the starting point Hamiltonian for DMRG or ED simulations (see \ref{sec:II} for details), where all parameters $U_i$ can be tuned freely. 

For QMC, to avoid sign problem explicitly, we need to rewrite the Hamiltonian in $\tau^\mu$ form by Fierz identity (see \ref{sec:III} for details)
\begin{equation}
H_{\tau} = \frac{1}{2}{\int{d\Omega_{1}{\int{d\Omega_{2}\delta\left( \left| {\Omega_{1} - \Omega_{2}} \right| \right){\sum\limits_{\mu = 0}^{3}{g_{\mu}\left( {\psi^{\dagger}\left( \Omega_{1} \right)\tau^{\mu}\psi\left( \Omega_{1} \right) - C\left( \Omega_{1} \right)\delta_{\mu,0}} \right)\left( {\psi^{\dagger}\left( \Omega_{2} \right)\tau^{\mu}\psi\left( \Omega_{2} \right) - C\left( \Omega_{2} \right)\delta_{\mu,0}} \right)}}}}}}
\end{equation}
with $C\left(\Omega_1\right)=2\sum_{m}\left|\Phi_m\left(\Omega_1\right)\right|^2$ ensure half-filling. According to $g_0=U_0+u_N,\ g_1=g_2=-(u_K+u_N),\ g_3=2u_N$, the sign-problem-free QMC simulation requires there are even negative terms (0 or 2) within $g_1,g_2,g_3$ and $g_0\geqslant0$ (see \ref{sec:III}). One can see this region covers the first quadrant of the phase diagram $(u_N,u_K)>0$ in Fig.~\ref{fig:fig1} of the main text. In our QMC simulation, we only focus on the $u_N=u_K$ SO(5) line. 

In our notation, the 10 generators of the SO(5) rotation group are
\begin{align}
L_{12} &= \tfrac{-i}{2}[\Gamma^1, \Gamma^2] = (\tau_x\otimes\mathbb{I})(\tau_y\otimes\mathbb{I}) - (\tau_y\otimes\mathbb{I})(\tau_x\otimes\mathbb{I}) = \tau_z \otimes\mathbb{I}, \\
L_{13} &= \tfrac{-i}{2}[\Gamma^1, \Gamma^3] = (\tau_x\otimes\mathbb{I})(\tau_z\otimes{\sigma_x}) - (\tau_z\otimes{\sigma_x})(\tau_x\otimes\mathbb{I}) = - \tau_y \otimes{\sigma_x}, \\
L_{14} &= \tfrac{-i}{2}[\Gamma^1, \Gamma^4] = (\tau_x\otimes\mathbb{I})(\tau_z\otimes{\sigma_y}) - (\tau_z\otimes{\sigma_y})(\tau_x\otimes\mathbb{I}) = - \tau_y \otimes{\sigma_y}, \\
L_{15} &= \tfrac{-i}{2}[\Gamma^1, \Gamma^5] = (\tau_x\otimes\mathbb{I})(\tau_z\otimes{\sigma_z}) - (\tau_z\otimes{\sigma_z})(\tau_x\otimes\mathbb{I}) = - \tau_y \otimes{\sigma_z}, \\
L_{23} &= \tfrac{-i}{2}[\Gamma^2, \Gamma^3] = (\tau_y\otimes\mathbb{I})(\tau_z\otimes{\sigma_x}) - (\tau_z\otimes{\sigma_x})(\tau_y\otimes\mathbb{I}) = \tau_x \otimes{\sigma_x}, \\
L_{24} &= \tfrac{-i}{2}[\Gamma^2, \Gamma^4] = (\tau_y\otimes\mathbb{I})(\tau_z\otimes{\sigma_y}) - (\tau_z\otimes{\sigma_y})(\tau_y\otimes\mathbb{I}) = \tau_x \otimes{\sigma_y}, \\
L_{25} &= \tfrac{-i}{2}[\Gamma^2, \Gamma^5] = (\tau_y\otimes\mathbb{I})(\tau_z\otimes{\sigma_z}) - (\tau_z\otimes{\sigma_z})(\tau_y\otimes\mathbb{I}) = \tau_x \otimes{\sigma_z}, \\
L_{34} &= \tfrac{-i}{2}[\Gamma^3, \Gamma^4] = (\tau_z\otimes{\sigma_x})(\tau_z\otimes{\sigma_y}) - (\tau_z\otimes{\sigma_y})(\tau_z\otimes{\sigma_x}) = \mathbb{I}\otimes{\sigma_z}, \\
L_{35} &= \tfrac{-i}{2}[\Gamma^3, \Gamma^5] = (\tau_z\otimes{\sigma_x})(\tau_z\otimes{\sigma_z}) - (\tau_z\otimes{\sigma_z})(\tau_z\otimes{\sigma_x}) = - \mathbb{I}\otimes{\sigma_y}, \\
L_{45} &= \tfrac{-i}{2}[\Gamma^4, \Gamma^5] = (\tau_z\otimes{\sigma_y})(\tau_z\otimes{\sigma_z}) - (\tau_z\otimes{\sigma_z})(\tau_z\otimes{\sigma_y}) = \mathbb{I}\otimes{\sigma_x}.
\end{align}

\subsection{Spherical Landau level}
For electrons moving on the surface of a sphere with $4\pi s$ monopole ($2s\in Z$), the Hamiltonian is $H_0 = \tfrac{1}{2M_er^2}\Lambda_\mu^2$, and $\Lambda_\mu=\partial_\mu + iA_\mu$. The eigenstates are quantized into spherical Landau levels with 
energies $E_n = [n(n+1)+(2n+1)s]/(2M_er^2)$ and $n=0,1,\cdots$ the Landau level index. The $(n+1)_\mathrm{th}$ level is $(2s+2n+1)$-fold degenerate. We assume all interactions are much smaller than the energy gap between Landau levels, and just consider the lowest Landau level (LLL) $n=0$, which is $(2s+1)$-fold degenerate and we denote $N=2s+1$ as the system size of the problem. The wave-functions of LLL orbital are monopole harmonics 
\begin{equation}
\Phi_m(\theta,\phi) = N_m e^{im\phi}\cos^{s+m}(\tfrac{\theta}{2})\sin^{s-m}(\tfrac{\theta}{2}),
\end{equation}
with $m=-s,-s+1,\cdots,s$ and $N_m = \sqrt{\tfrac{(2s+1)!}{4\pi (s+m)!(s-m)!}}$.

\subsection{Details on the LLL projection}
The projection of $H_\Gamma$ on the LLL of the Haldane sphere is carried out as
\begin{eqnarray}
H_\Gamma^{(LLL)} &=& \frac{1}{2} \int d\Omega_{1}\int d\Omega_{2}\delta\left( \left| {\Omega_{1} - \Omega_{2}} \right| \right)\sum\limits_{i = 0}^{5}U_{i}{\sum\limits_{m_{1},n_{1}}{\Phi_{m_{1}}^{*}\left( \Omega_{1} \right)\Phi_{n_{1}}\left( \Omega_{1} \right)\left( {{\sum\limits_{\alpha,\beta}{c_{m_{1},\alpha}^{\dagger}\Gamma_{\alpha,\beta}^{i}c_{n_{1},\beta}}} - 2\delta_{m_{1},n_{1}}\delta_{i,0}} \right)}} \nonumber\\
&&\cdot{\sum\limits_{m_{2},n_{2}}{\Phi_{m_{2}}^{*}\left( \Omega_{2} \right)\Phi_{n_{2}}\left( \Omega_{2} \right)\left( {{\sum\limits_{\alpha,\beta}{c_{m_{2},\alpha}^{\dagger}\Gamma_{\alpha,\beta}^{i}c_{n_{2},\beta}}} - 2\delta_{m_{2},n_{2}}\delta_{i,0}} \right)}},
\end{eqnarray}

and the projection of $H_\tau$ on the LLL of the Haldane sphere is carried out as
\begin{eqnarray}
H_{\tau}^{(LLL)} &=& \frac{1}{2} \int d\Omega_{1}\int d\Omega_{2}\delta\left( \left| {\Omega_{1} - \Omega_{2}} \right| \right)\sum\limits_{\mu = 0}^{3}g_{\mu}{\sum\limits_{m_{1},n_{1}}{\Phi_{m_{1}}^{*}\left( \Omega_{1} \right)\Phi_{n_{1}}\left( \Omega_{1} \right)\left( {{\sum\limits_{\alpha,\beta}{c_{m_{1},\alpha}^{\dagger}\tau_{\alpha,\beta}^{\mu}c_{n_{1},\beta}}} - 2\delta_{m_{1},n_{1}}\delta_{\mu,0}} \right)}} \nonumber\\
&&\cdot{\sum\limits_{m_{2},n_{2}}{\Phi_{m_{2}}^{*}\left( \Omega_{2} \right)\Phi_{n_{2}}\left( \Omega_{2} \right)\left( {{\sum\limits_{\alpha,\beta}{c_{m_{2},\alpha}^{\dagger}\tau_{\alpha,\beta}^{\mu}c_{n_{2},\beta}}} - 2\delta_{m_{2},n_{2}}\delta_{\mu,0}} \right)}}.
\end{eqnarray}
According to the Legendre polynomial $U\left( \left| {\mathbf{r}_{1} - \mathbf{r}_{2}} \right| \right) = {\sum\limits_{k = 0}^{\infty}{V_{k}P_{l}\left( {\cos\left( \Omega_{12} \right)} \right)}} = {\sum\limits_{k}{V_{k}\frac{4\pi}{2k+ 1}{\sum\limits_{m = - k}^{k}{Y_{k,m}^{*}\left( \Omega_{1} \right)Y_{k,m}\left( \Omega_{2} \right)}}}}$. For $U\left( \left| {\mathbf{r}_{1} - \mathbf{r}_{2}} \right| \right) = \delta\left( \left| {\Omega_{1} - \Omega_{2}} \right| \right)$, we have $V_k=2k+1$. 
We then arrive at the form, 
\begin{align}
H_\Gamma^{(LLL)} =& \sum_i U_i &\sum_{m_1,m_2,m}& (-1)^{2s+m+m_1+m_2}\tfrac{(2s+1)^2}{2}\sum_k V_k  
\begin{pmatrix}s&k&s\\-m_1&-m&m_1+m\end{pmatrix}
\begin{pmatrix}s&k&s\\-m_2&m&m_2-m\end{pmatrix}
\begin{pmatrix}s&k&s\\-s&0&s\end{pmatrix}^2 \nonumber\\
&\,&\times&  \left(c^\dag_{m_1,\alpha}\Gamma^i_{\alpha,\beta} c^{\,}_{m_1+m,\beta}-2\delta_{i0}\delta_{m0}\right)
\left(c^\dag_{m_2,\alpha}\Gamma^i_{\alpha,\beta} c^{\,}_{m_2-m,\beta}-2\delta_{i0}\delta_{m0}\right)\nonumber\\
=&\sum_i U_i &\sum_{m_1,m_2,m}& V_{m_1,m_2,m_2-m,m_1+m} \left(c^\dag_{m_1,\alpha}\Gamma^i_{\alpha,\beta} c^{\,}_{m_1+m,\beta}-2\delta_{i0}\delta_{m0}\right)
\left(c^\dag_{m_2,\alpha}\Gamma^i_{\alpha,\beta} c^{\,}_{m_2-m,\beta}-2\delta_{i0}\delta_{m0}\right)\label{HamG}
\end{align}
and 
\begin{align}
H_\tau^{(LLL)} =& \sum_\mu g_\mu &\sum_{m_1,m_2,m}& (-1)^{2s+m+m_1+m_2}\tfrac{(2s+1)^2}{2}\sum_k V_k  
\begin{pmatrix}s&k&s\\-m_1&-m&m_1+m\end{pmatrix}
\begin{pmatrix}s&k&s\\-m_2&m&m_2-m\end{pmatrix}
\begin{pmatrix}s&k&s\\-s&0&s\end{pmatrix}^2 \nonumber\\
&\,&\times&  \left(c^\dag_{m_1,\alpha}\tau^\mu_{\alpha,\beta} c^{\,}_{m_1+m,\beta}-2\delta_{\mu0}\delta_{m0}\right)
\left(c^\dag_{m_2,\alpha}\tau^\mu_{\alpha,\beta} c^{\,}_{m_2-m,\beta}-2\delta_{\mu0}\delta_{m0}\right)\nonumber\\
=&\sum_\mu g_\mu &\sum_{m_1,m_2,m}& V_{m_1,m_2,m_2-m,m_1+m} \left(c^\dag_{m_1,\alpha}\tau^\mu_{\alpha,\beta} c^{\,}_{m_1+m,\beta}-2\delta_{\mu0}\delta_{m0}\right)
\left(c^\dag_{m_2,\alpha}\tau^\mu_{\alpha,\beta} c^{\,}_{m_2-m,\beta}-2\delta_{\mu0}\delta_{m0}\right)\label{HamT}
\end{align}
with 
\begin{equation}
V_{m_1,m_2,m_3,m_4} = (-1)^{2s+m_1+2m_2-m_3}\tfrac{(2s+1)^2}{2} \sum_k (2k+1)  
\begin{pmatrix}s&k&s\\-m_1&m_1-m_4&m_4\end{pmatrix}
\begin{pmatrix}s&k&s\\-m_2&m_2-m_3&m_3\end{pmatrix}
\begin{pmatrix}s&k&s\\-s&0&s\end{pmatrix}^2. 
\end{equation}

\section{Detailed implementation in DMRG}
\label{sec:II}
\subsection{SU(2) symmetric Hamiltonian}
For the case of $u_1=u_2=u_K$ and $u_3=u_4=u_5=u_N$ considered in the main text, the model possesses the 
$\text{SU(2)}_\text{spin}\times\text{U(1)}_\text{valley}\times\text{U(1)}_\text{charge}$ symmetries. 
In this case, the projected SO(5) model Hamiltonian $H^{(LLL)}_{\Gamma}$ [c.f. \Eq{HamG}] can be rewritten into 
a spin rotation invariant and valley charge conserved form as 
\begin{align*}
H_\Gamma^{(LLL)} =&\sum_{m_1,m_2,m} &V_{m_1,m_2,m_2-m,m_1+m} &\sum_i U_i  \left(c^\dag_{m_1,\alpha}\Gamma^i_{\alpha,\beta} c^{\,}_{m_1+m,\beta}-2\delta_{i0}\delta_{m0}\right)
\left(c^\dag_{m_2,\alpha}\Gamma^i_{\alpha,\beta} c^{\,}_{m_2-m,\beta}-2\delta_{i0}\delta_{m0}\right)\\
=&\sum_{m_1,m_2,m}& V_{m_1,m_2,m_2-m,m_1+m} &~ \{ ~ U_0 \left(\Psi^\dag_{m_1} \Psi^{\,}_{m_1+m}-2\delta_{m0}\right)
\left(\Psi^\dag_{m_2} \Psi^{\,}_{m_2-m}-2\delta_{m0}\right)\\
&\,& \,- & ~\,~2u_K \left(\Psi^\dag_{m_1} \tau^+ \Psi^{\,}_{m_1+m}\right)
\left(\Psi^\dag_{m_2}\tau^- \Psi^{\,}_{m_2-m}\right)\\
&\,& \,- & ~\,~ 2u_K \left(\Psi^\dag_{m_1} \tau^- \Psi^{\,}_{m_1+m}\right)
\left(\Psi^\dag_{m_2}\tau^+ \Psi^{\,}_{m_2-m}\right)\\
&\,& \,- & ~\,~ 4u_N \left(\Psi^\dag_{m_1} \mathbf{S}^\dag \Psi^{\,}_{m_1+m}\right)\cdot
\left(\Psi^\dag_{m_2} \mathbf{S} \Psi^{\,}_{m_2-m}\right)~\},\\
\end{align*}
where the irreducible operator (irop) for fermion annihilation is
\begin{equation}
\hat\Psi^{S=\frac{1}{2},S_z} = 
\begin{pmatrix}
\hat\Psi_1 \\
\hat\Psi_2
\end{pmatrix}
\text{ with }
\hat\Psi_\tau^{S=\frac{1}{2},S_z} = 
\begin{pmatrix}
-\hat c^{\,}_{\downarrow,\tau} \\
\hat c^{\,}_{\uparrow,\tau}
\end{pmatrix},
\end{equation}
where the components $\hat\Psi_\tau^{\frac{1}{2},+\frac{1}{2}}=-c_{\downarrow,\tau}$ and 
$\hat\Psi_\tau^{\frac{1}{2},-\frac{1}{2}}=c_{\uparrow,\tau}$ 
transform as the irreducible representation (irep) under SU(2) spin rotation group
$|S=\frac{1}{2}; S_z\rangle$ with $S_z=+\frac{1}{2}, -\frac{1}{2}$, respectively. 
For this, the relative sign in the first component is important, and this fermion annihilation 
operator corresponds to the defining representation, i.e., $S=\frac{1}{2}$ for SU(2).
And we have
\begin{equation}
\tau^+ = 
\begin{pmatrix}0&0&1&0\\0&0&0&1\\0&0&0&0\\0&0&0&0\end{pmatrix},~ 
\tau^- = 
\begin{pmatrix}0&0&0&0\\0&0&0&0\\1&0&0&0\\0&1&0&0\end{pmatrix},~ 
\end{equation}
$\mathbf{S}\equiv(-\frac{1}{\sqrt{2}}S^+, S^z, \frac{1}{\sqrt{2}}S^-)^\mathrm{T}$ with
\begin{equation}
S^+ = 
\begin{pmatrix}0&1&0&0\\0&0&0&0\\0&0&0&1\\0&0&0&0\end{pmatrix},~ 
S^z = \frac{1}{2}
\begin{pmatrix}1&0&0&0\\0&-1&0&0\\0&0&1&0\\0&0&0&-1\end{pmatrix},~ 
S^- = 
\begin{pmatrix}0&0&0&0\\1&0&0&0\\0&0&0&0\\0&0&1&0\end{pmatrix},~ 
\end{equation}
and 
$\mathbf{S}^\dag\equiv(-\frac{1}{\sqrt{2}}S^-, S^z, \frac{1}{\sqrt{2}}S^+)$.

\subsection{Angular-momentum-space matrix product state}
We consider the many-body wavefucntion in the lowest Landau level basis with spin and valley degrees of freedom, 
\begin{equation}
|\psi\rangle = \sum_{\alpha_{-s}\cdots\alpha_m\cdots\alpha_s} A_{\alpha_{-s}\cdots\alpha_m\cdots\alpha_s} \bigotimes_{m=-s}^s|\alpha\rangle_{m},
\end{equation}
where $|\alpha_m\rangle$ spans a 16-dimension local Hilbert space, obtained by the 16 ways of filling in electrons within the 4 states 
$|\sigma,\tau\rangle\in\{|\uparrow,1\rangle,|\downarrow,1\rangle,|\uparrow,2\rangle,|\downarrow,2\rangle\}$ 
for each orbital $m\in\{-s, -s+1, \cdots, s-1, s\}$.

The matrix product state (MPS) ansatz for this particular case then expresses as, 
\begin{equation}
|\psi\rangle = \sum_{\alpha_{-s}\cdots\alpha_m\cdots\alpha_s}\sum_{\beta_{-s}\cdots\beta_m\cdots\beta_{s-1}} 
(A^{[-s]})_{\alpha_{-s}}^{\beta_{-s}} (A^{[-s+1]})_{\alpha_{-s+1}}^{\beta_{-s},\beta_{-s+1}}\cdots
(A^{[s]})_{\alpha_{-s+1}}^{\beta_{s-1}}
 \bigotimes_{m=-s}^s|\alpha\rangle_{m},
\end{equation}
where the geometric bond basis $|\beta_m\rangle$ is introduced to encode entanglement in the system. Graphically, it can be 
depicted as shown in \Fig{fig:figs1}.

\begin{figure}[h!]
	\includegraphics[width=0.55\columnwidth]{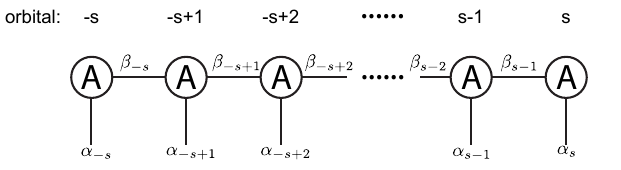}
	\caption{\textbf{Matrix product state (MPS) representation for the spherical Landau level systems.}
	}
	\label{fig:figs1}
\end{figure}

Practically, the 16-dimensional local Hilbert space is numerically costly for 2-site update in DMRG, and we thus split the two valleys into two adjacent 
local tensors and the more practical MPS reads
\begin{equation}
|\psi\rangle = \sum_{\alpha_{-s}^1\alpha_{-s}^2\cdots\alpha_s^1\alpha_s^2}\sum_{\beta_{-s}^1\beta_{-s}^2\cdots\beta_{s-1}^1\beta_{s-1}^2} 
\prod_{m=-s}^s
(A^{[m,1]})_{\alpha_m^1}^{\beta_{m-1}^2\beta_m^1}(A^{[m,2]})_{\alpha_m^2}^{\beta_m^1\beta_m^2}
 \bigotimes_{m=-s}^s|\alpha\rangle_{m}^1|\alpha\rangle_{m}^2,
\end{equation}
where the local Hilbert space $|\alpha_m^\tau\rangle$ now spans a 4-dimension local Hilbert space, 
obtained by the 4 ways of filling in the spin-up and/or spin-down electrons. 
The graphical representation is shown in \Fig{fig:figs2}.

\begin{figure}[h!]
	\includegraphics[width=0.55\columnwidth]{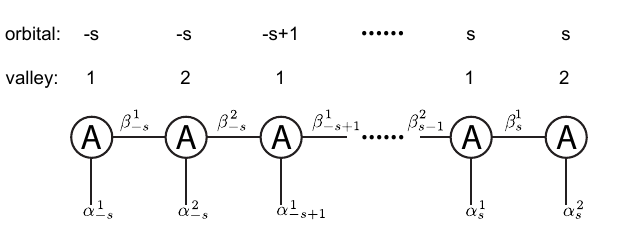}
	\caption{\textbf{Alternative matrix product state (MPS) representation for the spherical Landau level systems.}
	}
	\label{fig:figs2}
\end{figure}

\section{Detailed implementation in QMC}
\label{sec:III}
\subsection{Fierz identity}
Following the Ref.~\cite{impoliteHalf2018}, we use the Fierz identity to rewrite Hamiltonian from $H_\Gamma$ to $H_\tau$. 
First, we would like to introduce the property of $\Gamma^i=\left(\tau_x\bigotimes I_2,\ \tau_y\bigotimes I_2,\ \tau_z\bigotimes\sigma_x,\ \tau_z\bigotimes\sigma_y,\ \tau_z\bigotimes\sigma_z\right)$ matrices
\begin{equation}
\left( {\psi^{\dagger}\tau_{z}\psi} \right)^{2} = - \frac{1}{2}\left( {\psi^{\dagger}\psi - 2} \right)^{2} - \frac{1}{2}{\sum\limits_{i = 3,4,5}\left( {\psi^{\dagger}\Gamma^{i}\psi} \right)^{2}} + \frac{1}{2}{\sum\limits_{i = 1,2}\left( {\psi^{\dagger}\Gamma^{i}\psi} \right)^{2}} + 2.
\end{equation}
This equation comes from the idea that any $4\times4$ matrix can be expanded by 16 matrices $O^i\in\left\{\tau_\alpha\bigotimes\sigma_\beta\right\}$ and $\left\{O^i\otimes O^j\right\}$ forms the basis for $16\times16$ matrix.
\begin{equation}
O_{\alpha,\beta}^{i}O_{\gamma,\eta}^{i} = {\sum\limits_{j,k}{b_{i;j,k}O_{\alpha,\eta}^{j}O_{\gamma,\beta}^{k}}}.
\end{equation}
This formula times $O_{\beta,\gamma}^mO_{\eta,\alpha}^n$ and contract all labels
\begin{equation}
\Tr\left( {O^{i}O^{m}O^{i}O^{n}} \right) = {\sum\limits_{j,k}{b_{i;j,k}\Tr\left( {O^{j}O^{n}} \right)\Tr\left( {O^{k}O^{m}} \right)}}.
\end{equation}
Since $O^i$ and $O^j$ always commute or anti-commute and $\Tr\left(O^jO^n\right)=4\delta_{j,n}$, we have $b_{i;j,k}=\pm\frac{1}{4}\delta_{j,k}
,O_{\alpha,\beta}^iO_{\gamma,\eta}^i=\sum_{j}\pm\frac{1}{4}{O_{\alpha,\eta}^jO_{\gamma,\beta}^j}$. Here $+$ for commute and $-$ for anti-commute. With this relationship, we can derive
\begin{equation}\label{eq:Fierz}
\left( {\psi^{\dagger}O^{i}\psi} \right)^{2} = {\sum\limits_{j} \mp\frac{1}{4} {\left( {\psi^{\dagger}O^{j}\psi} \right)^{2} + \mu\psi^{\dagger}\psi}}.
\end{equation}
This can be seen from
\begin{equation}
\left( {\psi^{\dagger}O^{i}\psi} \right)^{2} = \psi_{\alpha}^{\dagger}O_{\alpha,\beta}^{i}\psi_{\beta}\psi_{\gamma}^{\dagger}O_{\gamma,\eta}^{i}\psi_{\eta} = 4\psi_{\alpha}^{\dagger}\psi_{\alpha}\delta_{i,0} + \psi_{\alpha}^{\dagger}\psi_{\alpha} - O_{\alpha,\beta}^{i}O_{\gamma,\eta}^{i}\psi_{\alpha}^{\dagger}\psi_{\eta}\psi_{\gamma}^{\dagger}\psi_{\beta} = {\sum\limits_{j}\mp \frac{1}{4}{\left( {\psi^{\dagger}O^{j}\psi} \right)^{2} + \mu\psi^{\dagger}\psi}}.
\label{eq:eqS33}
\end{equation}
The chemical potential tuning term is $5\psi^\dag\psi$ if $O^i=I_4$ and $\psi^\dag\psi$ if $O^i\neq I_4$. Directly expand the formula on the LHS below according to \Eq{eq:eqS33}, we obtain
\begin{equation}
\left( {\psi^{\dagger}\tau_{z}\psi} \right)^{2} - {\sum\limits_{i = 1,2}\left( {\psi^{\dagger}\Gamma^{i}\psi} \right)^{2}} + \left( {\psi^{\dagger}\psi} \right)^{2} = - \left( {\psi^{\dagger}\tau_{z}\psi} \right)^{2} - {\sum\limits_{i = 3,4,5}\left( {\psi^{\dagger}\Gamma^{i}\psi} \right)^{2}} + 4\psi^{\dagger}\psi.
\end{equation}
By using this formula, we can rewrite
\begin{equation} \label{eq:Htau}
g_{0}\left( {\psi^{\dagger}\psi - 2} \right)^{2} + g_{1}{\sum\limits_{\mu = x,y}\left( {\psi^{\dagger}\tau_{\mu}\psi} \right)^{2}} + g_{2}\left( {\psi^{\dagger}\tau_{z}\psi} \right)^{2} - 2g_{2} = \frac{U_0}{2}\left( {\psi^{\dagger}\psi - 2} \right)^{2} - \frac{u_{N}}{2}{\sum\limits_{i = 3,4,5}\left( {\psi^{\dagger}\Gamma^{i}\psi} \right)^{2}} - \frac{u_{K}}{2}{\sum\limits_{i = 1,2}\left( {\psi^{\dagger}\Gamma^{i}\psi} \right)^{2}},
\end{equation}
where the coefficients meet
\begin{equation}
g_{0} = \frac{U_0 + g_{2}}{2},g_{1} = - \frac{u_{K} + g_{2}}{2},g_{2} = u_{N}.
\end{equation}
Ignore the constant, we rewrite the Hamiltonian from the $\Gamma^i$ form to $\tau_\mu$ form, where $\sigma_\mu$ label is $2\times2$ identity matrix. This is crucial to have an explicit sign-problem-free determinant QMC simulation.

\subsection{Introducing the auxiliary field}
In the QMC, we first rewrite the Hamiltonian $H^{(LLL)}_{\tau}$ [c.f. \Eq{HamT}] in a compact form and integrate the solid angle based on spin-weighted spherical harmonics formula
\begin{equation}
H_{\tau}^{(LLL)} = \frac{1}{2}{\sum\limits_{\mu = 0}^{3}{\sum\limits_{l = 0}^{2s}{U_{\mu,l}{\sum\limits_{m = - l}^{l}{\delta\rho_{\mu,l,m}\delta\rho_{\mu,l,m}^{\dagger}}}}}} = \frac{1}{2}{\sum\limits_{\mu = 0}^{3}{\sum\limits_{l = 0}^{2s}{U_{\mu,l}\frac{1}{2}{\sum\limits_{m \geq 0}{\left( {\delta\rho_{\mu,l,m} + \delta\rho_{\mu,l,m}^{\dagger}} \right)^{2} - \left( {\delta\rho_{\mu,l,m} - \delta\rho_{\mu,l,m}^{\dagger}} \right)^{2}}}}}},
\end{equation}
where
\begin{eqnarray}
\delta\rho_{\mu,l,m} &=& {\sum\limits_{m_{1},n_{1}}{\sqrt{\frac{4\pi}{2l + 1}}{\int{d\Omega_{1}}}\Phi_{m_{1}}^{*}\left( \Omega_{1} \right)\Phi_{n_{1}}\left( \Omega_{1} \right)Y_{lm}^{*}\left( \Omega_{1} \right)\left( {{\sum\limits_{\alpha,\beta}{c_{m_{1},\alpha}^{\dagger}\tau_{\alpha,\beta}^{\mu}c_{n_{1},\beta}}} - 2\delta_{m_{1},n_{1}}\delta_{\mu,0}} \right)}} \nonumber\\
&=& {\sum\limits_{m_{1},n_{1}}{\left( {- 1} \right)^{s + n_{1}}\left( {2s + 1} \right)
		\begin{pmatrix}
		s & l & s \\
		{- m_{1}} & {- m} & n_{1} \\
		\end{pmatrix}\begin{pmatrix}
		s & l & s \\
		{- s} & 0 & s \\
		\end{pmatrix}\left({{\sum\limits_{\alpha,\beta}{c_{m_{1},\alpha}^{\dagger}\tau_{\alpha,\beta}^{\mu}c_{n_{1},\beta}}} - 2\delta_{m_{1},n_{1}}\delta_{\mu,0}} \right)}} \nonumber\\
&=& \left( {- 1} \right)^{m}\delta\rho_{\mu,l, - m}^{\dagger},
\end{eqnarray}
where $U_{\mu,l}=g_\mu\left(2l+1\right)$.
From now on, it is in a form that Hubbard–Stratonovich transformation can be carried out explicitly. We take the label $\mu$ at the outside because if there is no projection, terms with different $\mu$ commute and Trotter decomposition will not introduce error between different $\mu$ blocks. Then we would arrange $m$ label from large to small and then separate two auxiliary fields, since for $m=0$ and $m>s$, $\left[\delta\rho_{\mu,l,m}\pm\delta\rho_{\mu,l,m}^\dag,\delta\rho_{\mu,l^\prime,m}\pm\delta\rho_{\mu,l^\prime,m}^\dag\right]=0$. And we finally put $l$ label from large to small. The partition function of this Hamiltonian after Trotter decomposition and Hubbard–Stratonovich transformation is
\begin{eqnarray}
Z &=& \Tr\left( e^{- \beta H_{\tau}} \right) = \Tr\left( {\prod\limits_{t}e^{- \Delta t{\sum\limits_{\mu,m,l}{\frac{U_{\mu,l}}{4}{\lbrack{{({\delta\rho_{\mu,l,m} + \delta\rho_{\mu,l,m}^{\dagger}})}^{2} - {({\delta\rho_{\mu,l,m} - \delta\rho_{\mu,l,m}^{\dagger}})}^{2}}\rbrack}}}}} \right) \nonumber\\
&\approx& {\sum\limits_{\{ s_{t,\mu,l,m}\}}{{\prod\limits_{t,\mu,l,m}\left\lbrack {\frac{1}{16}\gamma\left( s_{t,\mu,l,m,1} \right)\gamma\left( s_{t,\mu,l,m,2} \right)} \right\rbrack}\Tr\left( {\prod\limits_{t,\mu,m}{{\prod\limits_{l}e^{i\eta{(s_{t,\mu,l,m,1})}A_{\mu,l}{({\delta\rho_{\mu,l,m} + \delta\rho_{\mu,l,m}^{\dagger}})}}}{\prod\limits_{l}e^{\eta{(s_{t,\mu,l,m,1})}A_{\mu,l}{({\delta\rho_{\mu,l,m} - \delta\rho_{\mu,l,m}^{\dagger}})}}}}} \right)}},\nonumber \\
\end{eqnarray}
where $\left\{s_{t,\mu,l,m}\right\}$ is the set of auxiliary field, $A_{\mu,l}=\sqrt{\frac{\Delta t U_{\mu,l}}{4}},\ \gamma\left(\pm1\right)=1+\frac{\sqrt6}{3},\gamma\left(\pm2\right)=1-\frac{\sqrt6}{3},\eta\left(\pm1\right)=\pm\sqrt{2\left(3-\sqrt6\right)},\eta\left(\pm2\right)=\pm\sqrt{2\left(3+\sqrt6\right)}$. Since we write down all the auxiliary fields, we would like to discuss the computation complexity for this angular momentum QMC method now. As one can count easily that the amount of auxiliary fields is $N_t N^2$, where $N_t$ is the number of trotter decomposition layers and $N=2s+1$ is the system size. One update for a non-local coupling auxiliary field costs $N^3$, so that the total cost will be $N_t N^5$ for a single sweep. Compared with Hubbard model $N_t N^3$~\cite{Hirsch1985} and cut off momentum space QMC $N_t N^4$~\cite{zhangMomentum2021,zhangPolynomial2023}, this angular momentum QMC bears heavier cost. Besides, different from Hubbard model, to control trotter error coming from $\left[\delta\rho_{\mu,l,m}\pm\delta\rho_{\mu,l,m}^\dag,\delta\rho_{\mu,l^\prime,m}\pm\delta\rho_{\mu,l^\prime,m}^\dag\right]\neq0$, one need a larger $N_t$ for a larger $N$ and a suitable arrangement for the position of auxiliary fields.

\subsection{Absence of Sign problem}
One should notice this $\tau_\mu$ form has no sign problem. Since we have not write the $\sigma$ explicitly in our Hamiltonian, there is an SU(2) symmetry for the decoupled Hamiltonian. We can use this spin-like freedom to make the block diagonalized matrices form complex conjugate. The trick can be seen by noticing the particle-hole transformation will transform $\delta\rho_{\mu,l,m}$ to $-\left(\delta\rho_{\mu,l,m}^\dag\right)^\ast$. If $U_{\mu,l}>0$, this means
\begin{eqnarray}
i\eta\left(s_{t,\mu,l,m,1}\right)A_{\mu,l}\left(\delta\rho_{\mu,l,m}+\delta\rho_{\mu,l,m}^\dag\right)\rightarrow-i\eta\left(s_{t,\mu,l,m,1}\right)A_{\mu,l}\left(\delta\rho_{\mu,l,m}^\ast+\left(\delta\rho_{\mu,l,m}^\dag\right)^\ast\right), \nonumber\\
\eta\left(s_{t,\mu,l,m,1}\right)A_{\mu,l}\left(\delta\rho_{\mu,l,m}-\delta\rho_{\mu,l,m}^\dag\right)\rightarrow\eta\left(s_{t,\mu,l,m,1}\right)A_{\mu,l}\left(\delta\rho_{\mu,l,m}^\ast-\left(\delta\rho_{\mu,l,m}^\dag\right)^\ast\right).
\end{eqnarray}
These are just what we want as the case $\mu=0,3$ when $U>0,U_0>0$. But one should notice for $\mu=1,2$, $U_{\mu,l}<0$ and we may need another minus sign for these terms and keep the formula at $\mu=0,3$ unchanged. This can be done by giving a minus sign phase for either $\tau_\pm$  particle (e.g., $c_{n_1,\tau_-}\rightarrow-c_{n_1,\tau_-}$ and $c_{n_1,\tau_-}^\dag\rightarrow-c_{n_1,\tau_-}^\dag$ ) because $\delta\rho_{\mu,l,m}$ is diagonal with $\tau$ at $\mu=0,3$ and off-diagonal with $\tau$ at $\mu=1,2$. From the discussion above, we can conclude one possible transformation for $\sigma_-$ particles
\begin{eqnarray}
{\tilde{c}}_{n_1,\tau_+,\sigma_-}=c_{n_1,\tau_+,\sigma_-}^\dag, \nonumber\\
{\tilde{c}}_{n_1,\tau_-,\sigma_-}=-c_{n_1,\tau_-,\sigma_-}^\dag.
\end{eqnarray}
As a check, with this transformation we explicitly have
\begin{eqnarray}
\delta\rho_{0,l,m,\sigma_{-}} &=& {\sum\limits_{m_{1},n_{1}}{\sqrt{\frac{4\pi}{2l + 1}}{\int{d\Omega_{1}}}\Phi_{m_{1}}^{*}\left( \Omega_{1} \right)\Phi_{n_{1}}\left( \Omega_{1} \right)Y_{lm}^{*}\left( \Omega_{1} \right)\left( {{\sum\limits_{\alpha = \pm 1}{c_{m_{1},\alpha,\sigma_{-}}^{\dagger}c_{n_{1},\alpha,\sigma_{-}}}} - \delta_{m_{1},n_{1}}} \right)}} \nonumber\\
&=& {\sum\limits_{m_{1},n_{1}}{\sqrt{\frac{4\pi}{2l + 1}}{\int{d\Omega_{1}}}\Phi_{m_{1}}^{*}\left( \Omega_{1} \right)\Phi_{n_{1}}\left( \Omega_{1} \right)Y_{lm}^{*}\left( \Omega_{1} \right)\left( {{\sum\limits_{\alpha = \pm 1}{\left( {- 1} \right)^{2\alpha}\tilde{c}_{m_{1},\alpha,\sigma_{-}}{\tilde{c}}_{n_{1},\alpha,\sigma_{-}}^{\dagger}}} - \delta_{m_{1},n_{1}}} \right)}} \nonumber\\
&=& - {\sum\limits_{m_{1},n_{1}}{\sqrt{\frac{4\pi}{2l + 1}}{\int{d\Omega_{1}}}\Phi_{m_{1}}^{*}\left( \Omega_{1} \right)\Phi_{n_{1}}\left( \Omega_{1} \right)Y_{lm}^{*}\left( \Omega_{1} \right)\left( {{\sum\limits_{\alpha = \pm 1}{{\tilde{c}}_{n_{1},\alpha,\sigma_{-}}^{\dagger}{\tilde{c}}_{m_{1},\alpha,\sigma_{-}}}} - \delta_{m_{1},n_{1}}} \right)}} \nonumber\\
&=& - {\sum\limits_{m_{1},n_{1}}{\sqrt{\frac{4\pi}{2l + 1}}{\int{d\Omega_{1}}}\Phi_{n_{1}}^{*}\left( \Omega_{1} \right)\Phi_{m_{1}}\left( \Omega_{1} \right)Y_{lm}^{*}\left( \Omega_{1} \right)\left( {{\sum\limits_{\alpha = \pm 1}{{\tilde{c}}_{m_{1},\alpha,\sigma_{-}}^{\dagger}{\tilde{c}}_{n_{1},\alpha,\sigma_{-}}}} - \delta_{m_{1},n_{1}}} \right)}} \nonumber\\
&=& - \delta\rho_{0,l, - m,\sigma_{+}}^{*}, \nonumber\\
\delta\rho_{1,l,m,\sigma_{-}} &=& {\sum\limits_{m_{1},n_{1}}{\sqrt{\frac{4\pi}{2l + 1}}{\int{d\Omega_{1}}}\Phi_{m_{1}}^{*}\left( \Omega_{1} \right)\Phi_{n_{1}}\left( \Omega_{1} \right)Y_{lm}^{*}\left( \Omega_{1} \right)\left( {c_{m_{1},\alpha,\sigma_{-}}^{\dagger}c_{n_{1}, - \alpha,\sigma_{-}} + c_{m_{1}, - \alpha,\sigma_{-}}^{\dagger}c_{n_{1},\alpha,\sigma_{-}}} \right)}} \nonumber\\
&=& {\sum\limits_{m_{1},n_{1}}{\sqrt{\frac{4\pi}{2l + 1}}{\int{d\Omega_{1}}}\Phi_{m_{1}}^{*}\left( \Omega_{1} \right)\Phi_{n_{1}}\left( \Omega_{1} \right)Y_{lm}^{*}\left( \Omega_{1} \right)\left( {- {\tilde{c}}_{m_{1},\alpha,\sigma_{-}}{\tilde{c}}_{n_{1}, - \alpha,\sigma_{-}}^{\dagger} - {\tilde{c}}_{m_{1}, - \alpha,\sigma_{-}}{\tilde{c}}_{n_{1},\alpha,\sigma_{-}}^{\dagger}} \right)}} \nonumber\\
&=& {\sum\limits_{m_{1},n_{1}}{\sqrt{\frac{4\pi}{2l + 1}}{\int{d\Omega_{1}}}\Phi_{m_{1}}^{*}\left( \Omega_{1} \right)\Phi_{n_{1}}\left( \Omega_{1} \right)Y_{lm}^{*}\left( \Omega_{1} \right)\left( {{\tilde{c}}_{n_{1}, - \alpha,\sigma_{-}}^{\dagger}{\tilde{c}}_{m_{1},\alpha,\sigma_{-}} + {\tilde{c}}_{n_{1},\alpha,\sigma_{-}}^{\dagger}{\tilde{c}}_{m_{1}, - \alpha,\sigma_{-}}} \right)}} \nonumber\\
&=& {\sum\limits_{m_{1},n_{1}}{\sqrt{\frac{4\pi}{2l + 1}}{\int{d\Omega_{1}}}\Phi_{n_{1}}^{*}\left( \Omega_{1} \right)\Phi_{m_{1}}\left( \Omega_{1} \right)Y_{lm}^{*}\left( \Omega_{1} \right)\left( {{\tilde{c}}_{m_{1},\alpha,\sigma_{-}}^{\dagger}{\tilde{c}}_{n_{1}, - \alpha,\sigma_{-}} + {\tilde{c}}_{m_{1}, - \alpha,\sigma_{-}}^{\dagger}{\tilde{c}}_{n_{1},\alpha,\sigma_{-}}} \right)}} \nonumber\\
&=& \delta\rho_{1,l, - m,\sigma_{+}}^{*}, \nonumber\\
\delta\rho_{2,l,m,\sigma_{-}} &=& - i{\sum\limits_{m_{1},n_{1}}{\sqrt{\frac{4\pi}{2l + 1}}{\int{d\Omega_{1}}}\Phi_{m_{1}}^{*}\left( \Omega_{1} \right)\Phi_{n_{1}}\left( \Omega_{1} \right)Y_{lm}^{*}\left( \Omega_{1} \right)\left( {c_{m_{1},\alpha,\sigma_{-}}^{\dagger}c_{n_{1}, - \alpha,\sigma_{-}} - c_{m_{1}, - \alpha,\sigma_{-}}^{\dagger}c_{n_{1},\alpha,\sigma_{-}}} \right)}}, \nonumber\\
&=& - i{\sum\limits_{m_{1},n_{1}}{\sqrt{\frac{4\pi}{2l + 1}}{\int{d\Omega_{1}}}\Phi_{m_{1}}^{*}\left( \Omega_{1} \right)\Phi_{n_{1}}\left( \Omega_{1} \right)Y_{lm}^{*}\left( \Omega_{1} \right)\left( {- {\tilde{c}}_{m_{1},\alpha,\sigma_{-}}{\tilde{c}}_{n_{1}, - \alpha,\sigma_{-}}^{\dagger} + {\tilde{c}}_{m_{1}, - \alpha,\sigma_{-}}{\tilde{c}}_{n_{1},\alpha,\sigma_{-}}^{\dagger}} \right)}} \nonumber\\
&=& - i{\sum\limits_{m_{1},n_{1}}{\sqrt{\frac{4\pi}{2l + 1}}{\int{d\Omega_{1}}}\Phi_{m_{1}}^{*}\left( \Omega_{1} \right)\Phi_{n_{1}}\left( \Omega_{1} \right)Y_{lm}^{*}\left( \Omega_{1} \right)\left( {{\tilde{c}}_{n_{1}, - \alpha,\sigma_{-}}^{\dagger}{\tilde{c}}_{m_{1},\alpha,\sigma_{-}} - {\tilde{c}}_{n_{1},\alpha,\sigma_{-}}^{\dagger}{\tilde{c}}_{m_{1}, - \alpha,\sigma_{-}}} \right)}} \nonumber\\
&=& i{\sum\limits_{m_{1},n_{1}}{\sqrt{\frac{4\pi}{2l + 1}}{\int{d\Omega_{1}}}\Phi_{n_{1}}^{*}\left( \Omega_{1} \right)\Phi_{m_{1}}\left( \Omega_{1} \right)Y_{lm}^{*}\left( \Omega_{1} \right)\left( {{\tilde{c}}_{m_{1},\alpha,\sigma_{-}}^{\dagger}{\tilde{c}}_{n_{1}, - \alpha,\sigma_{-}} - {\tilde{c}}_{m_{1}, - \alpha,\sigma_{-}}^{\dagger}{\tilde{c}}_{n_{1},\alpha,\sigma_{-}}} \right)}} \nonumber\\
&=& \delta\rho_{2,l, - m,\sigma_{+}}^{*}, \nonumber\\
\delta\rho_{3,l,m,\sigma_{-}} &=& {\sum\limits_{m_{1},n_{1}}{\sqrt{\frac{4\pi}{2l + 1}}{\int{d\Omega_{1}}}\Phi_{m_{1}}^{*}\left( \Omega_{1} \right)\Phi_{n_{1}}\left( \Omega_{1} \right)Y_{lm}^{*}\left( \Omega_{1} \right)\left( {c_{m_{1},\alpha,\sigma_{-}}^{\dagger}c_{n_{1},\alpha,\sigma_{-}} - c_{m_{1}, - \alpha,\sigma_{-}}^{\dagger}c_{n_{1}, - \alpha,\sigma_{-}}} \right)}} \nonumber\\
&=& {\sum\limits_{m_{1},n_{1}}{\sqrt{\frac{4\pi}{2l + 1}}{\int{d\Omega_{1}}}\Phi_{m_{1}}^{*}\left( \Omega_{1} \right)\Phi_{n_{1}}\left( \Omega_{1} \right)Y_{lm}^{*}\left( \Omega_{1} \right)\left( {{\tilde{c}}_{m_{1},\alpha,\sigma_{-}}{\tilde{c}}_{n_{1},\alpha,\sigma_{-}}^{\dagger} - {\tilde{c}}_{m_{1}, - \alpha,\sigma_{-}}{\tilde{c}}_{n_{1}, - \alpha,\sigma_{-}}^{\dagger}} \right)}} \nonumber\\
&=& - {\sum\limits_{m_{1},n_{1}}{\sqrt{\frac{4\pi}{2l + 1}}{\int{d\Omega_{1}}}\Phi_{m_{1}}^{*}\left( \Omega_{1} \right)\Phi_{n_{1}}\left( \Omega_{1} \right)Y_{lm}^{*}\left( \Omega_{1} \right)\left( {{\tilde{c}}_{n_{1},\alpha,\sigma_{-}}^{\dagger}{\tilde{c}}_{m_{1},\alpha,\sigma_{-}} - {\tilde{c}}_{n_{1}, - \alpha,\sigma_{-}}^{\dagger}{\tilde{c}}_{m_{1}, - \alpha,\sigma_{-}}} \right)}} \nonumber\\
&=& - {\sum\limits_{m_{1},n_{1}}{\sqrt{\frac{4\pi}{2l + 1}}{\int{d\Omega_{1}}}\Phi_{n_{1}}^{*}\left( \Omega_{1} \right)\Phi_{m_{1}}\left( \Omega_{1} \right)Y_{lm}^{*}\left( \Omega_{1} \right)\left( {{\tilde{c}}_{m_{1},\alpha,\sigma_{-}}^{\dagger}{\tilde{c}}_{n_{1},\alpha,\sigma_{-}} - {\tilde{c}}_{m_{1}, - \alpha,\sigma_{-}}^{\dagger}{\tilde{c}}_{n_{1}, - \alpha,\sigma_{-}}} \right)}} \nonumber\\
&=& - \delta\rho_{3,l, - m,\sigma_{+}}^{*}.
\end{eqnarray}

With these relationship the nontrivial part contributing the sample weight has
\begin{eqnarray}
&&\Tr\left( {\prod\limits_{t,\mu,m}{{\prod\limits_{l}e^{i\eta{(s_{t,\mu,l,m,1})}A_{\mu,l}{({\delta\rho_{\mu,l,m,\sigma_{-}} + \delta\rho_{\mu,l,m,\sigma_{-}}^{\dagger}})}}}{\prod\limits_{l}e^{\eta{(s_{t,\mu,l,m,1})}A_{\mu,l}{({\delta\rho_{\mu,l,m,\sigma_{-}} - \delta\rho_{\mu,l,m,\sigma_{-}}^{\dagger}})}}}}} \right) \nonumber\\
&=& \Tr\left( {\prod\limits_{t,\mu,m}{{\prod\limits_{l}e^{i\eta{(s_{t,\mu,l,m,1})}A_{\mu,l}{({\delta\rho_{\mu,l,m,\sigma_{+}} + \delta\rho_{\mu,l,m,\sigma_{+}}^{\dagger}})}}}{\prod\limits_{l}e^{\eta{(s_{t,\mu,l,m,1})}A_{\mu,l}{({\delta\rho_{\mu,l,m,\sigma_{+}} - \delta\rho_{\mu,l,m,\sigma_{+}}^{\dagger}})}}}}} \right)^{*}
\end{eqnarray}
And this closes the proof of this section.

\subsection{Details for the ED and QMC measurements}
In ED simulation, the Hamiltonian will be block diagonalized according to good quantum number (i.e., particle number, total magnetic quantum number and total angular momentum quantum number). We diagonalize total angular momentum operator $J^2$ within the subspace with a certain particle number and total magnetic quantum number. Since the many-body Hamiltonian has nothing to do with the total magnetic quantum number, we just take the smallest total magnetic quantum number subspace to derive all the possible eigenvalues. We use the same unitary transformation diagonalizing total angular momentum $J^2$ to block diagonalize the Hamiltonian. Then the eigenstates within each block corresponds to the same total angular momentum quantum number. $J^2$ is defined as below
\begin{equation}
J^{2} = \left( {\sum\limits_{m,\alpha}J_{m,\alpha}} \right)^{2} = {\sum\limits_{m,n,\alpha,\beta}{J_{m,\alpha} \cdot J_{n,\beta}}} = {\sum\limits_{m,n,\alpha,\beta}{J_{m,\alpha}^{z}J_{n,\beta}^{z} + \frac{1}{2}\left( {J_{m,\alpha}^{+}J_{n,\beta}^{-} + J_{m,\alpha}^{-}J_{n,\beta}^{+}} \right)}},
\end{equation}
where
\begin{eqnarray}
J_{m,\alpha}^z&=&mc_{m,\alpha}^\dag c_{m,\alpha} \nonumber\\
J_{m,\alpha}^+&=&\sqrt{\left(s-m\right)\left(s+m+1\right)}c_{m+1,\alpha}^\dag c_{m,\alpha} \nonumber\\
J_{m,\alpha}^-&=&\sqrt{\left(s+m\right)\left(s-m+1\right)}c_{m-1,\alpha}^\dag c_{m,\alpha}
\end{eqnarray}
and $J_{-s,\alpha}^-=J_{s,\alpha}^+=0$. Insert the formula above to expression of $J^2$ will give us the total angular momentum operator in Fock basis. It is easy to verify $J_\alpha^z=\sum_{m} J_{m,\alpha}^z, J_\alpha^+=\sum_{m} J_{m,\alpha}^+$ and $J_\alpha^-=\sum_{m} J_{m,\alpha}^-$ do form angular momentum algebra
\begin{eqnarray}
\left\lbrack {J_{\alpha}^{+},J_{\alpha^{'}}^{-}} \right\rbrack &=& 2J_{\alpha}^{z}\delta_{\alpha,\alpha^{'}}, \nonumber\\
\left\lbrack {J_{\alpha}^{z},J_{\alpha^{'}}^{+}} \right\rbrack &=& J_{\alpha}^{+}\delta_{\alpha,\alpha^{'}}, \nonumber\\
\left\lbrack {J_{\alpha}^{z},J_{\alpha^{'}}^{-}} \right\rbrack &=& - J_{\alpha}^{-}\delta_{\alpha,\alpha^{'}}.
\end{eqnarray}
Besides, this definition also will not introduce minus sign from fermion anti-commutation if we just align the nearest magnetic quantum number orbital states together (i.e., $-s,-s+1,\ldots,s$) in each subspace $\alpha$. The measurements for ED is straightforward, and one need to expand multi-fermion correlation at a certain auxiliary field configuration by Wick's theorem.

We list the measurements in our ED and QMC simulation below. For the SO(5) order parameter, we define
\begin{eqnarray}
O_{i,l,m} &=& {\int{d\Omega_{1}Y_{lm}^{*}\left( \Omega_{1} \right)\psi^{\dagger}\left( \Omega_{1} \right)\Gamma^{i}\psi\left( \Omega_{1} \right)}} \nonumber\\
&=& {\sum\limits_{m^{'},n^{'}}{\int{d\Omega_{1}\Phi_{m^{'}}^{*}\left( \Omega_{1} \right)\Phi_{n^{'}}\left( \Omega_{1} \right)Y_{lm}^{*}\left( \Omega_{1} \right)c_{m^{'}}^{\dagger}\Gamma^{i}c_{n^{'}}}}} \nonumber\\
&=& \sqrt{\frac{2l + 1}{4\pi}}{\sum\limits_{m^{'},n^{'}}{\left( {- 1} \right)^{s + n^{'}}\left( {2s + 1} \right)
		\begin{pmatrix}
		s & l & s \\
		{- m^{'}} & {- m} & n^{'} \\
		\end{pmatrix}\begin{pmatrix}
		s & l & s \\
		{- s} & 0 & s \\
		\end{pmatrix}c_{m^{'}}^{\dagger}\Gamma^{i}c_{n^{'}}}} \nonumber\\
&\equiv& \sqrt{\frac{2l + 1}{4\pi}}{\sum\limits_{m^{'},n^{'}}{M_{m^{'},n^{'}}^{l,m}c_{m^{'}}^{\dagger}\Gamma^{i}c_{n^{'}}}}.
\end{eqnarray}
Then the imaginary time correlation function can be defined as
\begin{eqnarray}
&&\left\langle {O_{i,l,m}(t) O_{i,l,m}^{\dagger}(0)} \right\rangle \nonumber\\
&=& \frac{2l + 1}{4\pi}\left\langle {\left( {\sum\limits_{m_{1},n_{1}}{M_{m_{1},n_{1}}^{l,m}c_{m_{1}}^{\dagger}(t)\Gamma^{i}c_{n_{1}}(t)}} \right)\left( {\sum\limits_{m_{2},n_{2}}{\left( M_{m_{2},n_{2}}^{l,m} \right)^{*}c_{n_{2}}^{\dagger}(0)\Gamma^{i}c_{m_{2}}(0)}} \right)} \right\rangle \nonumber\\
&\equiv& {\sum\limits_{m_{1},n_{1},m_{2},n_{2}}{P_{m_{1},n_{1},m_{2},n_{2}}^{l,m}\left\langle {\left( {\sum\limits_{m_{1},n_{1}}{c_{m_{1}}^{\dagger}(t)\Gamma^{i}c_{n_{1}}(t)}} \right)\left( {\sum\limits_{m_{2},n_{2}}{c_{n_{2}}^{\dagger}(0)\Gamma^{i}c_{m_{2}}(0)}} \right)} \right\rangle}}.
\end{eqnarray}
Besides, we also use internal energy to benchmark our QMC simulation with ED 
\begin{eqnarray}
\left\langle H \right\rangle = {\sum\limits_{\mu,m_{1},n_{1},m_{2},n_{2}}V_{\mu,m_{1},n_{1},m_{2},n_{2}}}\left\langle {\left( {{\sum\limits_{\alpha_{1},\beta_{1}}{c_{m_{1},\alpha_{1}}^{\dagger}\tau_{\alpha_{1},\beta_{1}}^{\mu}c_{n_{1},\beta_{1}}}} - 2\delta_{m_{1},n_{1}}\delta_{\mu,0}} \right)\left( {{\sum\limits_{\alpha_{2},\beta_{2}}{c_{n_{2},\alpha_{2}}^{\dagger}\tau_{\alpha_{2},\beta_{2}}^{\mu}c_{m_{2},\beta_{2}}}} - 2\delta_{m_{2},n_{2}}\delta_{\mu,0}} \right)} \right\rangle, \nonumber\\
\end{eqnarray}
here $V_{\mu,m_{1},n_{1},m_{2},n_{2}} \equiv \frac{1}{2} \sum\limits_{l = 0}^{2s}U_{\mu,l}\sum\limits_{m = - l}^{l}M_{m_{1},n_{1}}^{l,m}\left( M_{m_{2},n_{2}}^{l,m} \right)^{*}$.

The observed results are listed in Figs.~\ref{fig:figS3}, \ref{fig:figS4} and \ref{fig:figS6}.

\begin{figure}[h!]
	\includegraphics[width=0.8\columnwidth]{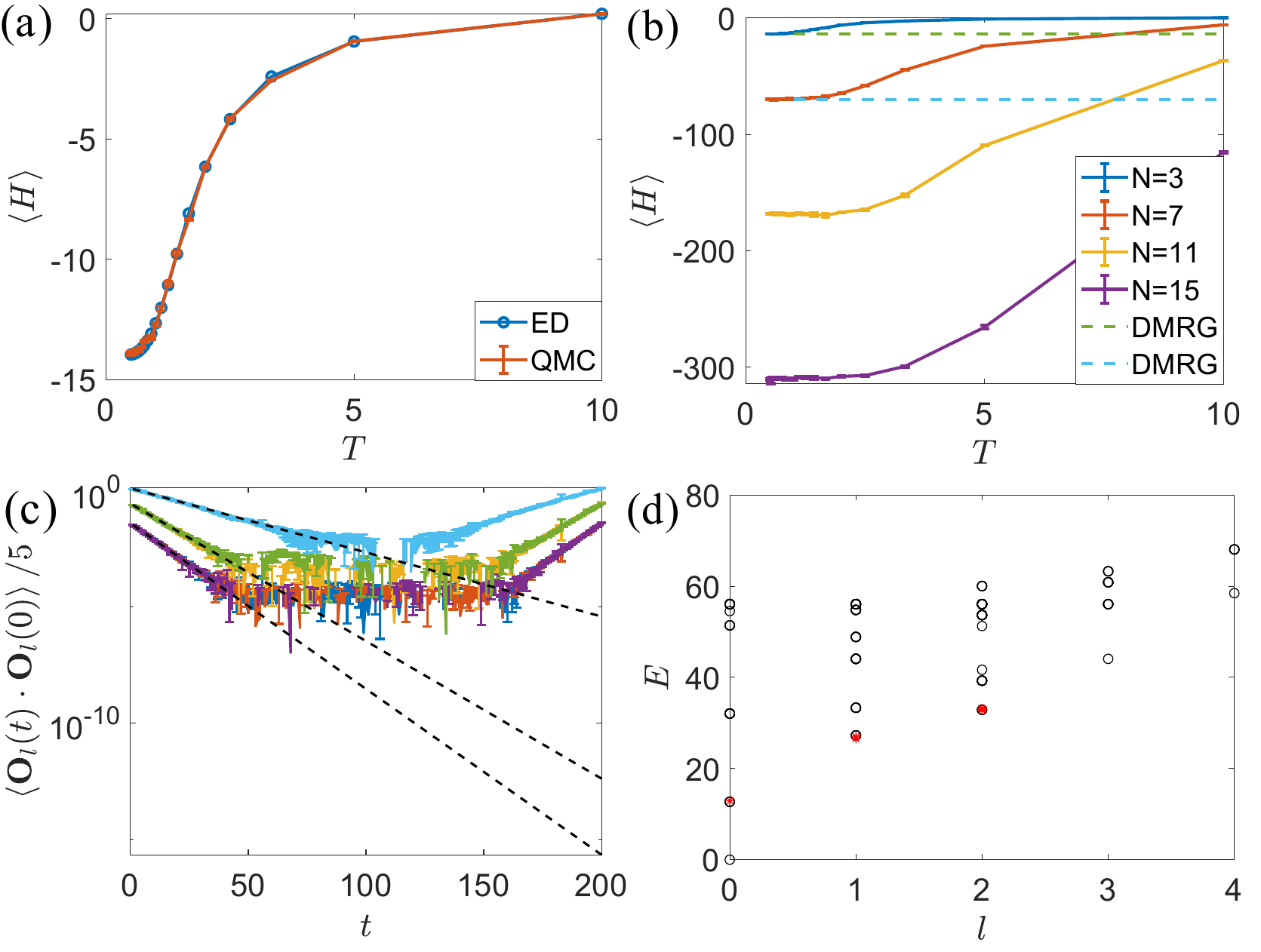}
	\caption{\textbf{Benchmark results from ED, QMC and DMRG at $u_N=u_K=U_0=1$}
		(a) Benchmark for internal energy from ED and QMC with $N=3$.
		(b) Benchmark for internal energy from finite temperature QMC and zero temperature DMRG.
		(c) Extract excitation gap from the imaginary time correlation of order parameter 
		$\langle\mathbf{O}_l(t)\cdot\mathbf{O}_l(0)\rangle/5$ for size $N=3$. The 3 groups of data correspond to $\mathbf{O}_{l=0}, \mathbf{O}_{l=1}, \mathbf{O}_{l=2}$ from gentle slope to steep slope. $\Delta t=0.01$ and the result is simulated at temperature $T=1/(200\Delta t)=0.5$ which is low enough. The dashed lines show the exponential fitting of $e^{-Et}$ where $E$ is the excitation gap.
		(d) Energy spectrum from ED and QMC excitation gap extracted from (c). The red stars are QMC results for $\mathbf{O}_{l=0}, \mathbf{O}_{l=1}, \mathbf{O}_{l=2}$ gaps and match well with ED $N=3$ lowest three charge neutral excitations. 
	}
	\label{fig:figS3}
\end{figure}

\begin{figure}[h!]
	\includegraphics[width=0.8\columnwidth]{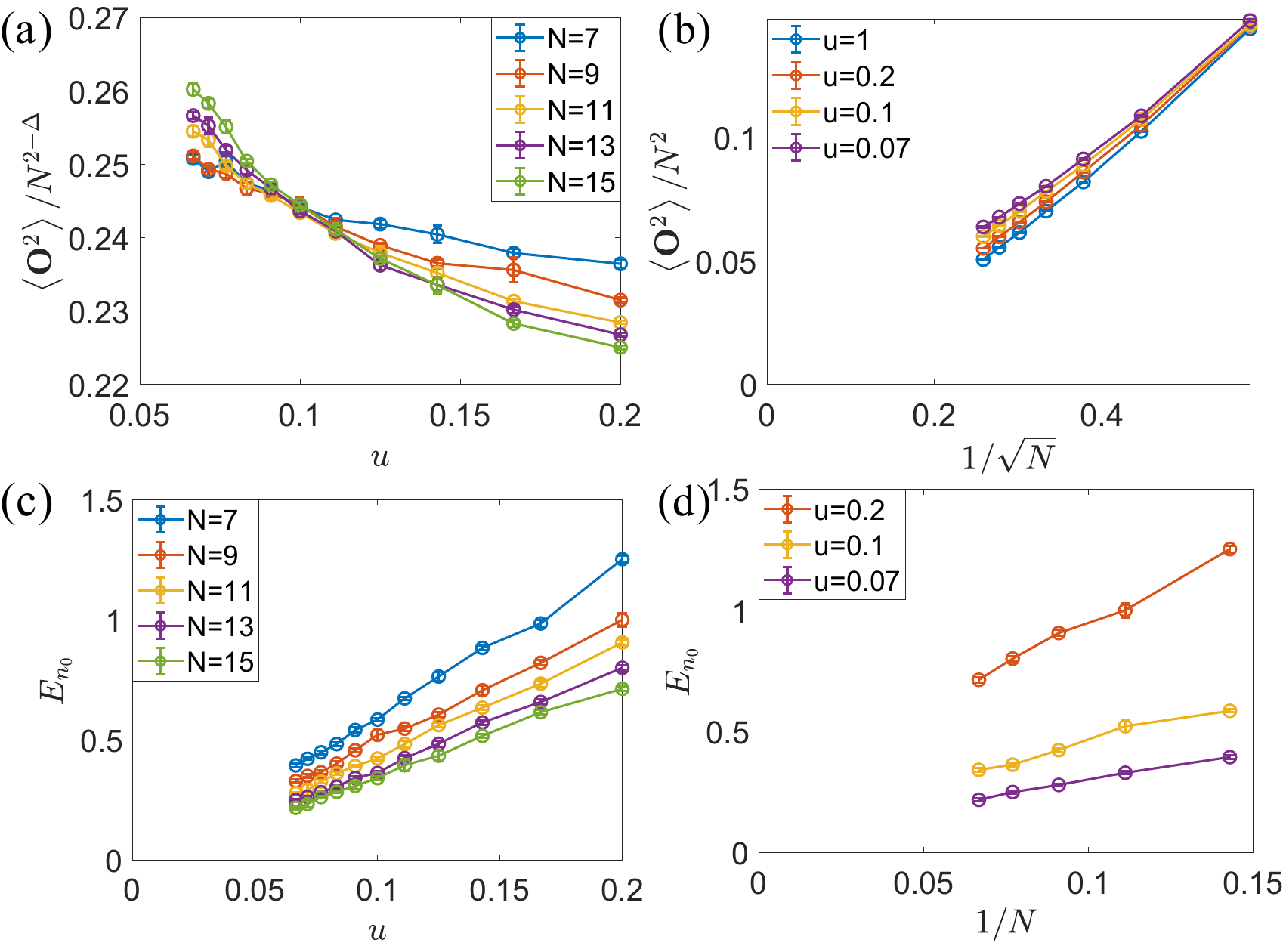}
	\caption{\textbf{QMC results for $0.07<u<1$ region}
		(a) Order parameter finite size crossing by assuming $\Delta=0.519$.
		(b) Order parameter for different u, indicating transition from order to disorder.
		(c) Excitation gaps extracted from the slope of $\left\langle O_{l=0}(t) O_{l=0}(0)\right\rangle $.
		(d) Extrapolation for excitation gaps. 
	}
	\label{fig:figS4}
\end{figure}

\begin{figure}[h!]
	\includegraphics[width=\columnwidth]{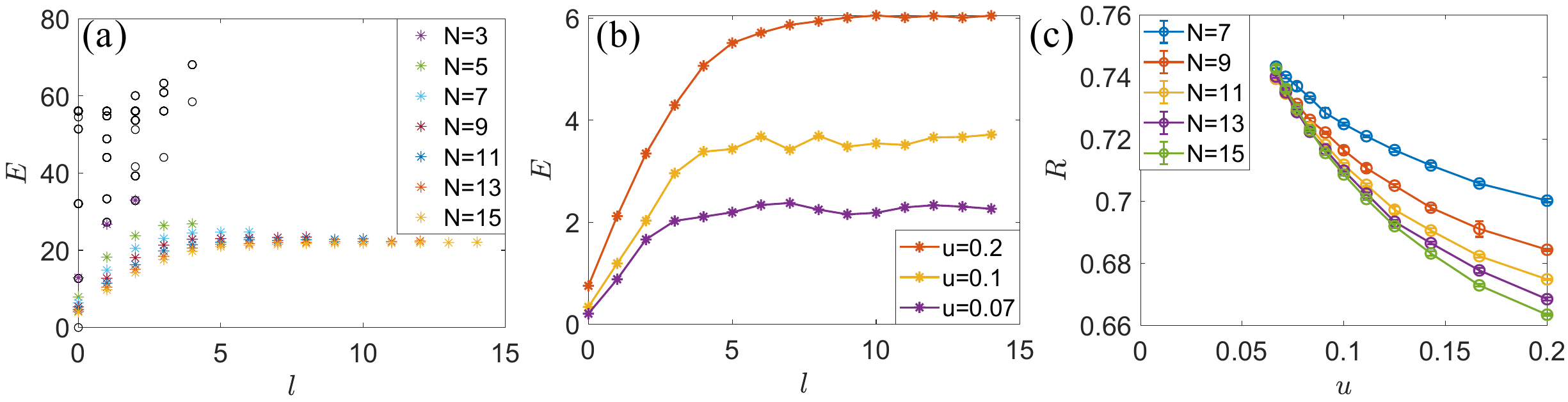}
	\caption{\textbf{QMC results for angular momentum resolution}
		(a) Black dots are from $N=3,u=1$ ED and colorful stars are extracted from the slope of $\left\langle O_{l}(t) O_{l}(0)\right\rangle$ in QMC simulation with size $N=3,5,7,9,11,13,15$. From smaller size to larger size, one has more angular momentum points and the dispersion for small $l$ converges.
		(b) Dispersion for different $u$ extracted from QMC with size $N=15$. One can see the dispersion becomes smoother when decrease $u$.
		(c) Correlation ratio defined as $R\equiv1-\left\langle O_{l=1}^2\right\rangle/\left\langle O_{l=0}^2\right\rangle$. Similar with the one defined in momentum space, the cross behavior of $R$ roughly determine the phase transition point $u\simeq0.1$.
	}
	\label{fig:figS6}
\end{figure}

\section{Crossing Point Analysis}
\label{sec:iv}
In this section, we provide the detailed derivation for the scaling form of 
the crossing points, such that the position of the critical point and the associated critical exponents can be obtained in a controlled manner from the finite size data. Such crossing point analysis has been widely applied and tested for quantum criticality of 2D Ising, SU(2) and other spin models~\cite{qinDuality2017,shaoQuantum2016,maAnomalous2018,maRole2019} and can be further traced back to Fisher's ``phenomenological renormalization'', which was first numerically tested with transfermatrix results for the Ising model in Ref.~\cite{luckCorrections1985}. We note, this is the first time, that such systematically analysis is applied to the DMRG data of quantum criticality, therefore our attempts also contribute to the methodological development for the DMRG investigations of quantum criticality.

Let's consider the standard form of finite-size scaling 
for an arbitrary observable, 
\begin{equation}
O(\delta, L) = L^{-\kappa/\nu} f(\delta L^{1/\nu}, \lambda L^{-\omega}).
\end{equation}
Here, $\delta=q-q_c$ is the deviation from the transition point $q_c$, and we also 
consider the correction from the leading irrelevant field $\lambda$ and its 
corresponding exponent $\omega$. 
In practise, due to the limit of computational resources, we only increase the system size
by $x$, and consider the crossing point of observable between size pair $(N, N+x)$. 
For the sake of notation simplicity, we express the scaling form as a function of total number of 
size $N$ instead of linear size $L\sim\sqrt{N}$, i.e., 
\begin{equation}
O(\delta, N) = N^{-\frac{\kappa}{2\nu}} f(\delta N^{\frac{1}{2\nu}}, \lambda N^{-\frac{\omega}{2}}) 
= N^{-\frac{\kappa}{2\nu}} (a_0 + a_1 \delta N^{\frac{1}{2\nu}} + b_1 N^{-\frac{\omega}{2}} + \cdots), 
\end{equation}
where the second equality relation is simply from Taylor's expansion up to first order. 
Similarly, for system size $N+x$, we have 
\begin{equation}
O(\delta, N+x) 
= (N+x)^{-\frac{\kappa}{2\nu}} (a_0 + a_1 \delta (N+x)^{\frac{1}{2\nu}} 
+ b_1 (N+x)^{-\frac{\omega}{2}} + \cdots). 
\end{equation}
Then, at the crossing point $\delta^\ast$, by definition we have 
$O(\delta^\ast,N) = O(\delta^\ast,N+x)$, which leads to the scaling form 
for the crossing point itself and the observable at the crossing point, 
\begin{equation}
\delta^\ast(N) = \frac{a_0}{a_1} \frac{(1+x/N)^{-\frac{\kappa}{2\nu}}-1}{1-(1+x/N)^{\frac{1-\kappa}{2\nu}}} N^{-\frac{1}{2\nu}} + \frac{b_1}{a_1} \frac{(1+x/N)^{-\frac{\omega}{2}-\frac{\kappa}{2\nu}}-1}{1-(1+x/N)^{\frac{1-\kappa}{2\nu}}} N^{-\frac{1}{2\nu}-\frac{\omega}{2}} + \cdots,
\end{equation}
and
\begin{equation}
O(\delta^\ast,N) = 
N^{-\frac{\kappa}{2\nu}}\left\{a_0 + a_1
\left[\frac{a_0}{a_1} \frac{(1+x/N)^{-\frac{\kappa}{2\nu}}-1}{1-(1+x/N)^{\frac{1-\kappa}{2\nu}}} N^{-\frac{1}{2\nu}} + \frac{b_1}{a_1} \frac{(1+x/N)^{-\frac{\omega}{2}-\frac{\kappa}{2\nu}}-1}{1-(1+x/N)^{\frac{1-\kappa}{2\nu}}} N^{-\frac{1}{2\nu}-\frac{\omega}{2}} + \cdots\right] N^{\frac{1}{2\nu}} + 
b_1 N^{-\frac{\omega}{2}} +\cdots \right\}.
\end{equation}

In the case of Binder ratio, we have $\kappa=0$ and when $x\ll N$, we then arrive at 
\begin{equation}\label{Eq:CrossingScaling}
\delta^\ast(N) = a N^{-\frac{1}{2\nu}-\frac{\omega}{2}} + \cdots ,
\end{equation}
and
\begin{equation}\label{Eq:BinderScaling}
U(\delta^\ast, N) = b + c N^{-\frac{\omega}{2}} + \cdots.
\end{equation}
From these two scaling forms, in principle, we can fit the finite-size data of 
Binder ratio and their crossing point, from which the correlation length 
exponent $\nu$ and $\omega$ can be obtained, and the thermodynamic limit value of the position of the critical point $\delta^*(N\to\infty)$ and the universal Binder ratio $U(\delta^*,N\to\infty)$. In the main text, we applied this method to determine the transition point between the SO(2) symmetry-breaking VBS phase and the SO(3) symmetry-breaking N\'eel phase with the disorder phase.
  
In practise, one can also fit $\omega$ from \Eq{Eq:BinderScaling} first, 
and then fit $\nu$ with fixed $\omega$ from \Eq{Eq:CrossingScaling}. 
The second fitting of $\nu$ depends on the accuracy of $\omega$ obtained from the 
first fitting, which may results in large uncertainty of $\nu$. 

To independently determine $\nu$, we can consider 
$U(\delta, N) = a_0 + a_1\delta N^{\frac{1}{2\nu}} + b_1 N^{-\frac{\omega}{2}} + 
c_1 \delta N^{\frac{1}{2\nu}-\frac{\omega}{2}}+\cdots$, whose 
first-order derivative with respect to $\delta$ writes 
\begin{equation}
U'(\delta, N) = a_1 N^{\frac{1}{2\nu}} + c_1 N^{\frac{1}{2\nu}-\frac{\omega}{2}} + \cdots.
\end{equation}
Then the difference of the logarithmic of the above equation between size pair 
$(N,N+x)$ will be 
\begin{equation}
\frac{2N}{x}\ln{\frac{U'(\delta^\ast,N+x)}{U'(\delta^\ast,N)}} = \frac{1}{\nu} - c N^{-\frac{\omega}{2}}.
\end{equation}
We define the finite-size value of $1/\nu$ as 
\begin{equation}
\frac{1}{\nu^\ast}(\delta^\ast,N) = \frac{2N}{x}\ln{\frac{U'(\delta^\ast,N+x)}{U'(\delta^\ast,N)}}, 
\end{equation}
and then have the finite-size scaling form for $1/\nu$, i.e., 
\begin{equation}
\frac{1}{\nu^\ast}(\delta^\ast,N)= \frac{1}{\nu} - c N^{-\frac{\omega}{2}}.
\label{Eq:nuScaling}
\end{equation}
This will provide an independent check for the validity of the $\nu$ obtained from \Eq{Eq:CrossingScaling}.

Overall, we make use of the \Eq{Eq:CrossingScaling}, \Eq{Eq:BinderScaling} and \Eq{Eq:nuScaling} upon the Binder ratio data, to independently and unbiasedly obtain the 
$\delta^*(N\to\infty)$, $\omega$ and $\nu$. These data are shown in Fig. 2 (a), (b), (c) in the main text for the VBS-Disorder transition at $u_K=2$, and in Figs.~\ref{fig:figCPAuk4}, ~\ref{fig:figDCuk4} for the VBS-Disorder transition at $u_K=4$, Figs.~\ref{fig:figCPAuk05}, ~\ref{fig:figDCuk05} for the VBS-Disorder transition at $u_K=0.5$, and Figs.~\ref{fig:figCPAun2}, ~\ref{fig:figDCun2} for the N\'eel-Disroder transition at $u_N=2$.

\begin{figure}[h!]
	\includegraphics[width=.55\columnwidth]{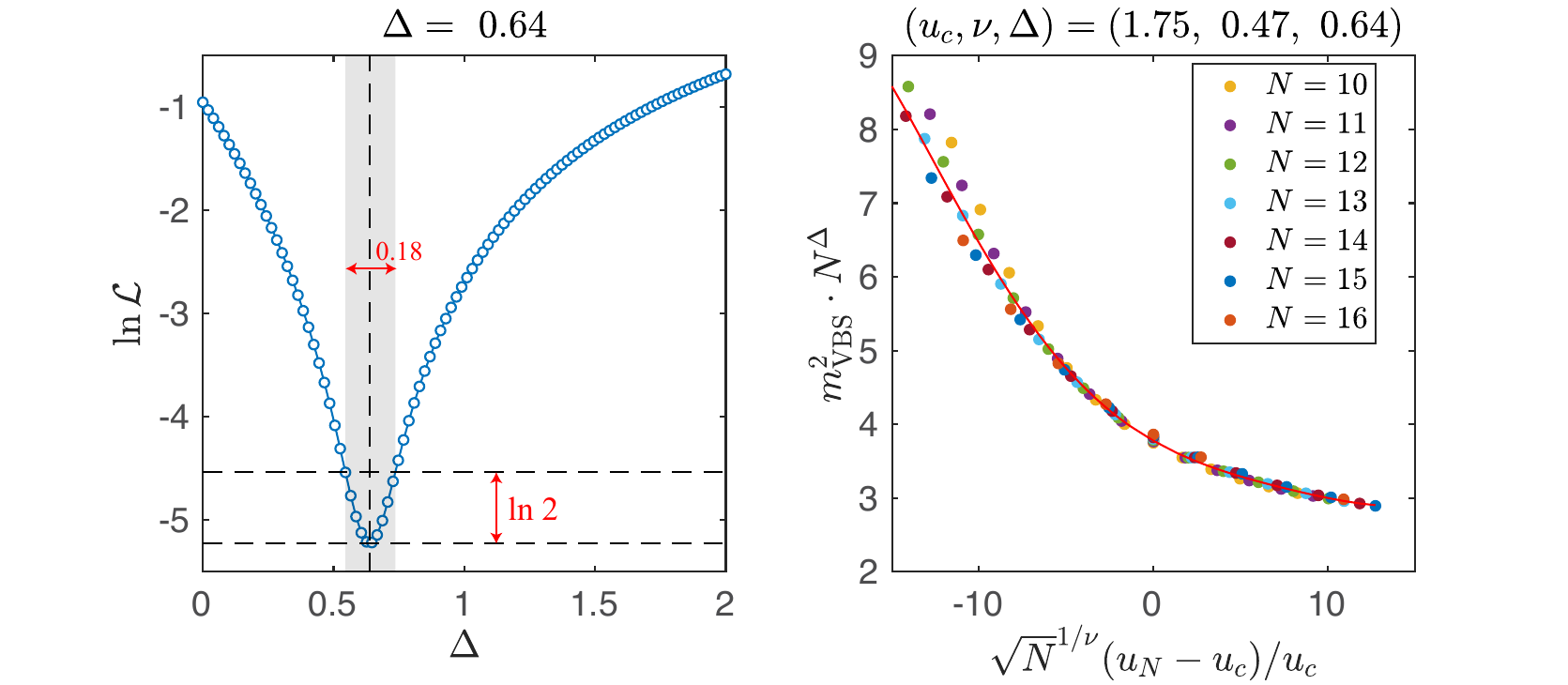}
	\caption{\textbf{More DMRG results for crossing point analysis.}\
Along the fixed $u_K=2$ cut, \\
(Left) Natural logarithmic of loss function $\mathcal{L}$ (defined as the squared deviation of the fitted scaling function away from the data points) versus $\Delta$, from which a optimized 
$\Delta=0.64(9)$ is obtained, and the error bar is defined as the grey range where the loss function is within 2 times of the optimal one.
\\
(Right) $m^2_\mathrm{VBS}$ rescaled by $N^{\Delta}$ with scaling dimension $\Delta\simeq0.64$ 
versus $\sqrt{N}^{1/\nu} (u_N-u_c)/u_c$, collapses nicely for various system sizes $N=10,9,...,16$.
}
\label{fig:figDCuk2}
\end{figure}

\begin{figure}[h!]
	\includegraphics[width=\columnwidth]{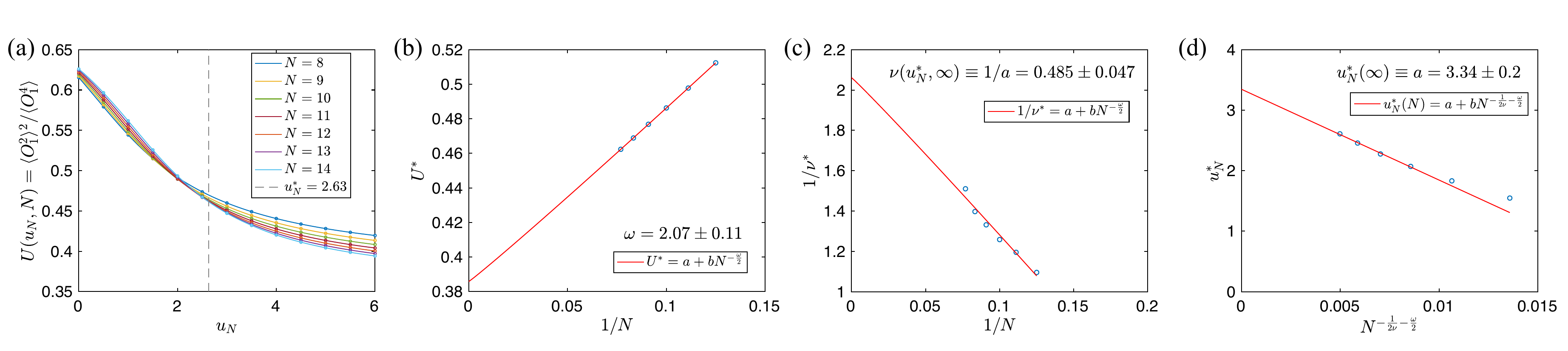}
	\caption{\textbf{More DMRG results for crossing point analysis.}\
Along the fixed $u_K=4$ cut, \\
(a) The VBS Binder ratio $U\equiv\langle O^2_1\rangle^2/\langle O^4_1\rangle$ crosses between successive size pair $(N,N+1)$, whose crossing points $u_N^\ast$ drift towards larger $u_N$ with larger $N$. \\
(b) The subleading operator exponent $\omega$ is obtained from the scaling form of Binder ratios value at crossing point, i.e. $U(u_N^\ast, N) = a + bN^{-\frac{\omega}{2}}$,
from which $\omega=2.1(1)$ is extracted. \\
(c) The correlation length exponent $\nu$ is obtained from the scaling form of the 
first-order derivatives of Binder ratios at crossing point, to be specific, 
$1/\nu^\ast(u_N^\ast,N) = 1/\nu + bN^{-\frac{\omega}{2}}$, from which $\nu=0.49(5)$ is 
extracted. \\
(d) The crossing point $u_N^\ast$'s are extrapolated to 
$u_c=3.3(2)$ in the thermodynamic limit with the scaling form $u_N^\ast(N) = u_c + N^{-\frac{1}{2\nu}-\frac{\omega}{2}}$, with $\nu=0.49(3)$ and $\omega=1.8(2)$.
}
\label{fig:figCPAuk4}
\end{figure}

\begin{figure}[h!]
	\includegraphics[width=.55\columnwidth]{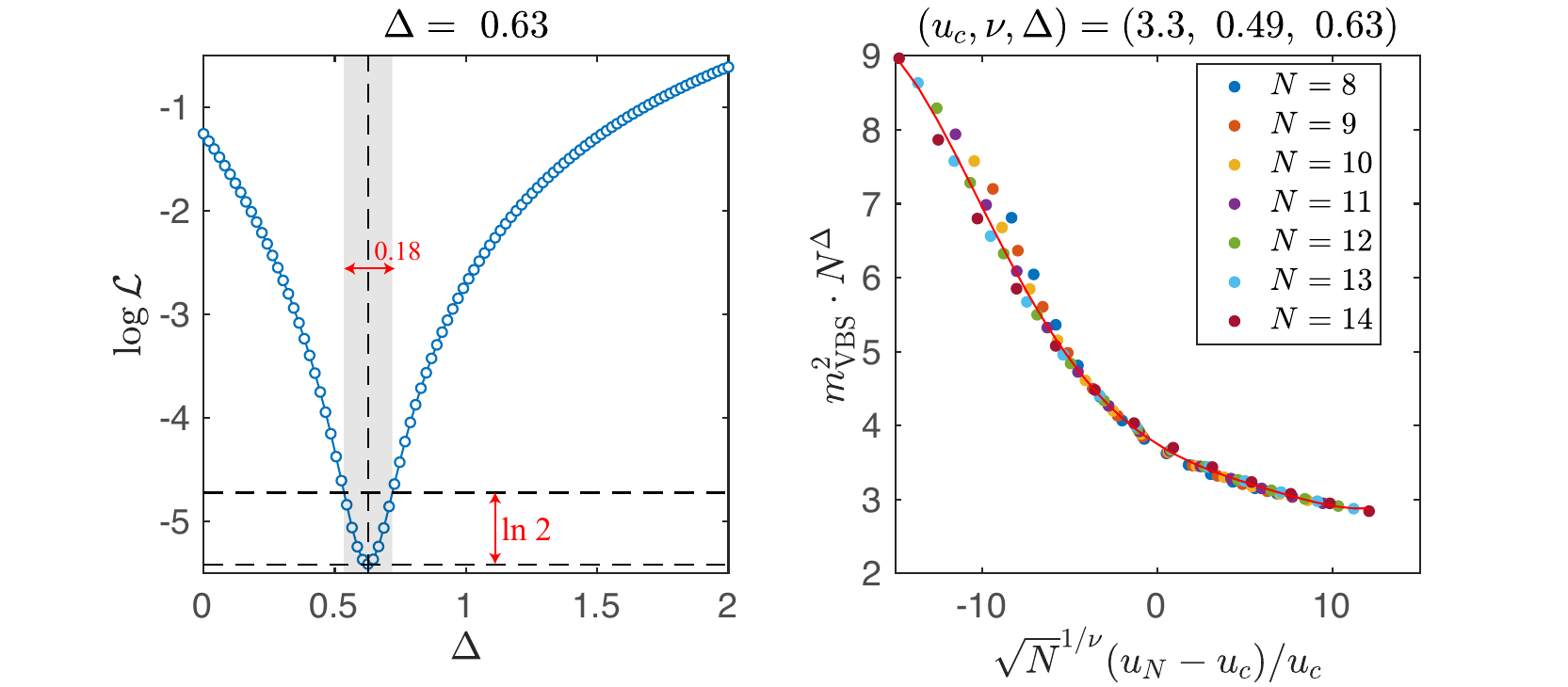}
	\caption{\textbf{More DMRG results for crossing point analysis.}\
Along the fixed $u_K=4$ cut, \\
(Left) Natural logarithmic of loss function $\mathcal{L}$ (defined as the squared deviation of the fitted scaling function away from the data points) versus $\Delta$, from which a optimized 
$\Delta=0.63(9)$ is obtained, and the error bar is defined as the grey range where the loss function is within 2 times of the optimal one.
\\
(Right) $m^2_\mathrm{VBS}$ rescaled by $N^{\Delta}$ with scaling dimension $\Delta\simeq0.63$ 
versus $\sqrt{N}^{1/\nu} (u_N-u_c)/u_c$, collapses nicely for various system sizes $N=8,9,...,14$.
}
\label{fig:figDCuk4}
\end{figure}

\begin{figure}[h!]
	\includegraphics[width=\columnwidth]{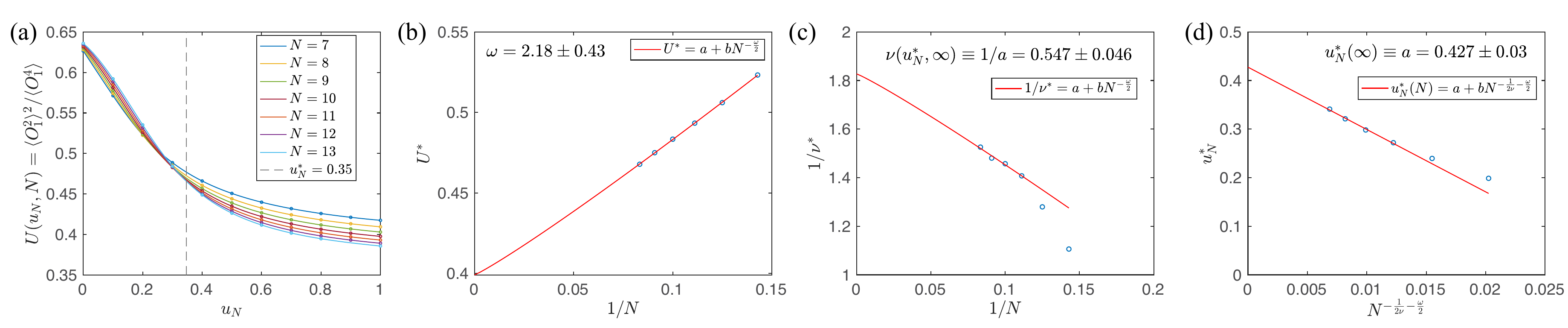}
	\caption{\textbf{More DMRG results for crossing point analysis.}\
Along the fixed $u_K=0.5$ cut, \\
(a) The VBS Binder ratio $U\equiv\langle O^2_1\rangle^2/\langle O^4_1\rangle$ crosses between successive size pair $(N,N+1)$, whose crossing points $u_N^\ast$ drift towards larger $u_N$ with larger $N$. \\
(b) The subleading operator exponent $\omega$ is obtained from the scaling form of Binder ratios value at crossing point, i.e. $U(u_N^\ast, N) = a + bN^{-\frac{\omega}{2}}$,
from which $\omega=2.2(4)$ is extracted. \\
(c) The correlation length exponent $\nu$ is obtained from the scaling form of the 
first-order derivatives of Binder ratios at crossing point, to be specific, 
$1/\nu^\ast(u_N^\ast,N) = 1/\nu + bN^{-\frac{\omega}{2}}$, from which $\nu=0.55(5)$ is 
extracted. \\
(d) The crossing point $u_N^\ast$'s are extrapolated to 
$u_c=0.43(3)$ in the thermodynamic limit with the scaling form $u_N^\ast(N) = u_c + N^{-\frac{1}{2\nu}-\frac{\omega}{2}}$, with $\nu=0.55(5)$ and $\omega=2.2(4)$.
}
\label{fig:figCPAuk05}
\end{figure}

\begin{figure}[h!]
	\includegraphics[width=.55\columnwidth]{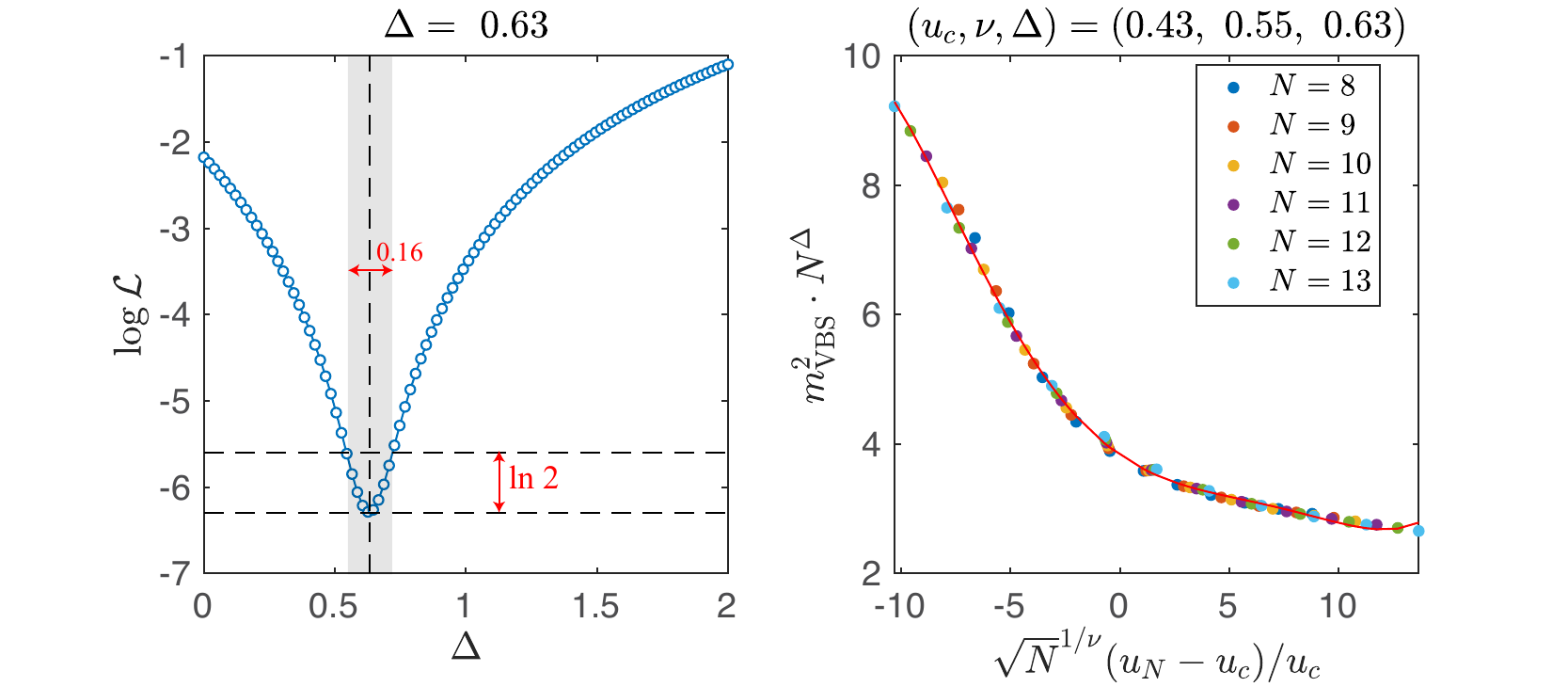}
	\caption{\textbf{More DMRG results for crossing point analysis.}\
	Along the fixed $u_K=0.5$ cut, \\
(Left) Natural logarithmic of loss function $\mathcal{L}$ (defined as the squared deviation of the fitted scaling function away from the data points) versus $\Delta$, from which a optimized 
$\Delta=0.63(8)$ is obtained, and the error bar is defined as the grey range where the loss function is within 2 times of the optimal one.
\\
(Right) $m^2_\mathrm{VBS}$ rescaled by $N^{\Delta}$ with scaling dimension $\Delta\simeq0.63$ 
versus $\sqrt{N}^{1/\nu} (u_N-u_c)/u_c$, collapses nicely for various system sizes $N=8,9,...,13$.
}
\label{fig:figDCuk05}
\end{figure}

\begin{figure}[h!]
	\includegraphics[width=\columnwidth]{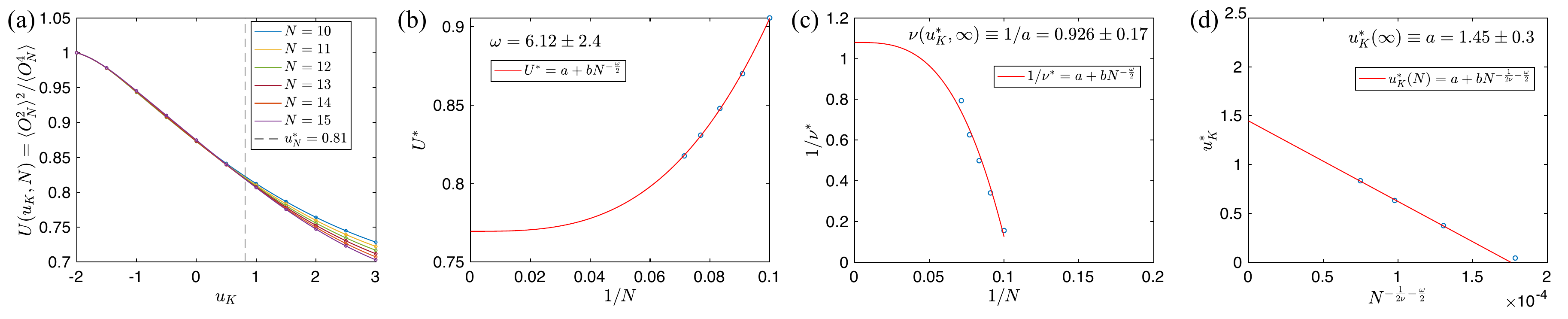}
	\caption{\textbf{More DMRG results for crossing point analysis.}\
Along the fixed $u_N=2$ cut, \\
(a) The N\'eel Binder ratio $U_\text{N\'eel}\equiv\langle O^2_N\rangle^2/\langle O^4_N\rangle$ crosses between successive size pair $(N,N+1)$, whose crossing points $u_K^\ast$ drift towards larger $u_K$ with larger $N$. \\
(b) The subleading operator exponent $\omega$ is obtained from the scaling form of Binder ratios value at crossing point, i.e. $U(u_K^\ast, N) = a + bN^{-\frac{\omega}{2}}$,
from which $\omega=6(2)$ is extracted. \\
(c) The correlation length exponent $\nu$ is obtained from the scaling form of the 
first-order derivatives of Binder ratios at crossing point, to be specific, 
$1/\nu^\ast(u_K^\ast,N) = 1/\nu + bN^{-\frac{\omega}{2}}$, from which $\nu=0.9(2)$ is 
extracted. \\
(d) The crossing point $u_K^\ast$'s are extrapolated to 
$u_c=1.5(3)$ in the thermodynamic limit with the scaling form $u_K^\ast(N) = u_c + N^{-\frac{1}{2\nu}-\frac{\omega}{2}}$, with $\nu=0.9(2)$ and $\omega=6(2)$.
}
\label{fig:figCPAun2}
\end{figure}

\begin{figure}[h!]
	\includegraphics[width=.55\columnwidth]{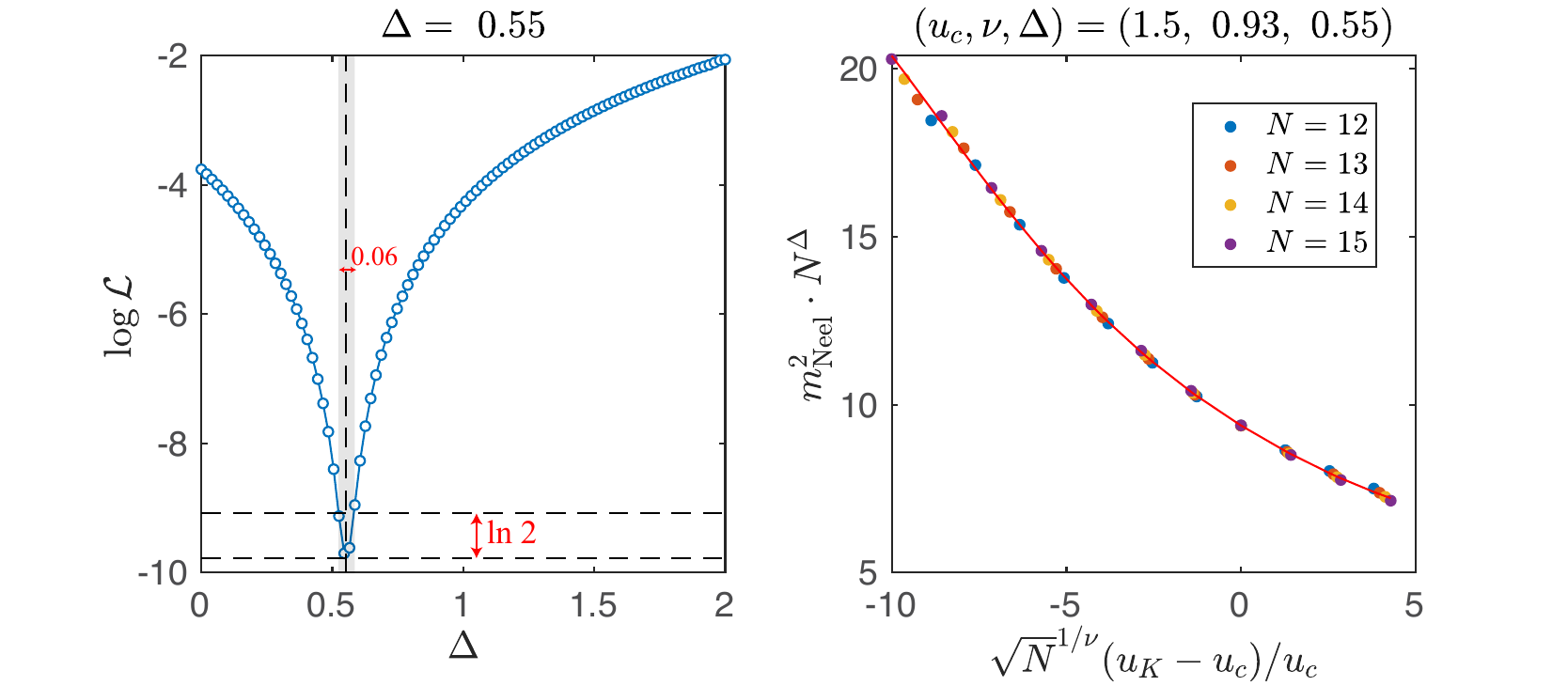}
	\caption{\textbf{More DMRG results for crossing point analysis.}\
Along the fixed $u_N=2$ cut, \\
(Left) Natural logarithmic of loss function $\mathcal{L}$ (defined as the squared deviation of the fitted scaling function away from the data points) versus $\Delta$, from which a optimized 
$\Delta=0.55(3)$ is obtained, and the error bar is defined as the grey range where the loss function is within 2 times of the optimal one.
\\
(Right) $m^2_\text{N\'eel}$ rescaled by $N^{\Delta}$ with scaling dimension $\Delta\simeq0.55$ 
versus $\sqrt{N}^{1/\nu} (u_K-u_c)/u_c$, collapses nicely for various system sizes $N=12,13,14,15$.
}
\label{fig:figDCun2}
\end{figure}

{
\section{More DMRG results for the phase separation region }
\label{sec:v}

In this section, in order to distinguish the two ``disordered'' phases in our SO(5) phase diagram,
we've calculated a few of lowest-lying states in different sectors with fixed particle number $N_e$.
As shown in \Fig{fig:figPS} (a), in the case of $u_K=u_N=u=-4$, 
we can see that, after doping 4 extra particles/holes,
the energies are lower than that in the half-filled sectors (indicated by the grey dashed line).
This is different from the right plot ($u=4$), where the half-filling energy is the lowest one.
This is the evidence that the ``disordered'' phase in the 3rd quadrants of the phase diagram is
actually phase separation consist of the entirely empty and entirely filled cases.

\begin{figure}[h!]
	\includegraphics[width=.75\columnwidth]{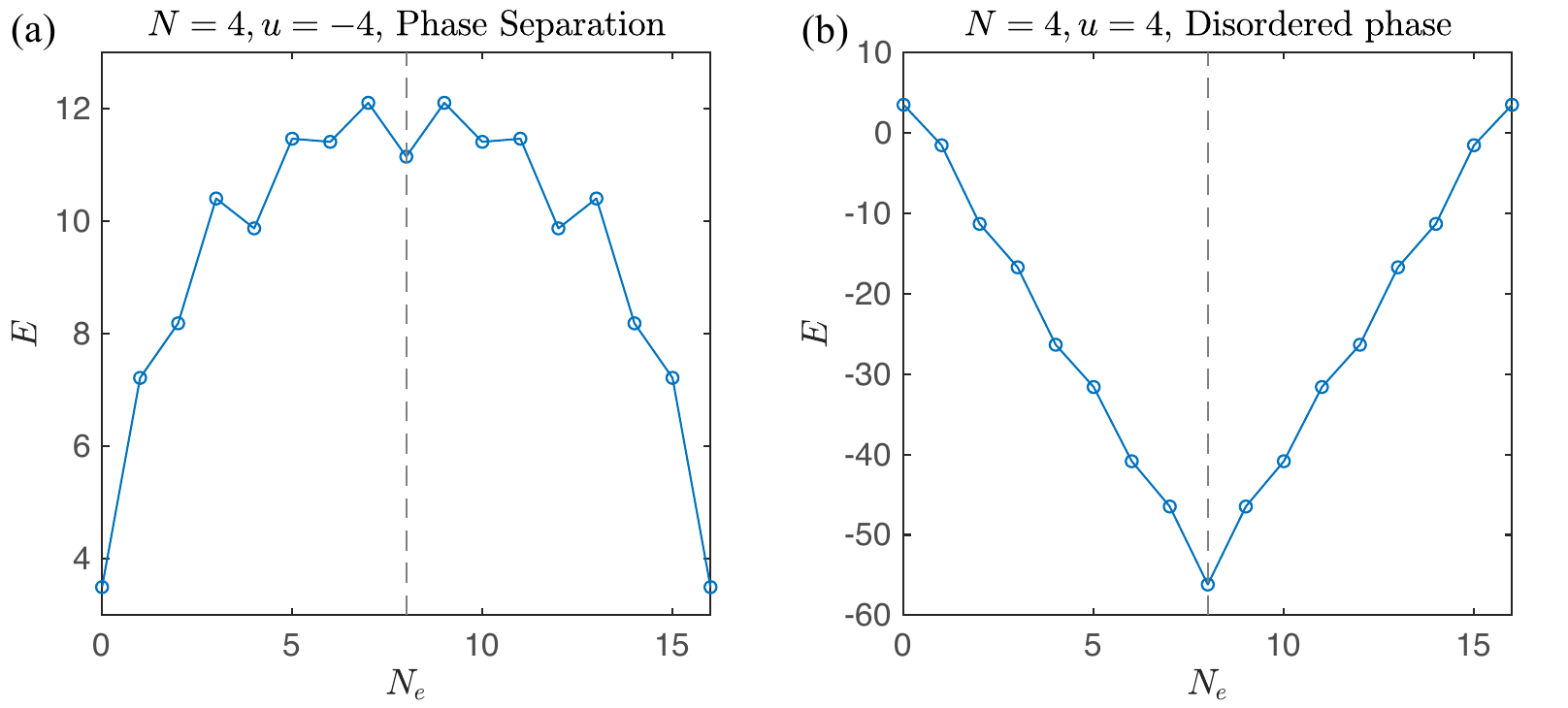}
	\caption{\textbf{DMRG results for the lowest-lying states in each fixed-particle-number sector.}\\
(a) For the case of $u_K=u_N=u=-4$, the lowest energies in different sectors with fixed particle number 
$N_e$ is calculated, where the dashed line indicates the half-filled sector.\\
(b) For the case of $u_K=u_N=u=4$, the lowest energies in different sectors with fixed particle number 
$N_e$ is calculated, where the dashed line indicates the half-filled sector.\\
}
\label{fig:figPS}
\end{figure}
}

\section{More DMRG results for the first-order transitions}
\label{sec:vi}

{
In this section, we will study the first-order transitions in the quadrants other than the first quadrant. 

Specifically, for the valley polarization (VP) to  VBS transition,
the phase boundary can be seen from the 
$\tau_\mu$ form \Eq{eq:Htau}  of the Hamiltonian derived in \Sec{sec:III}, i.e., 
$$
H = g_{0}\left( {\psi^{\dagger}\psi - 2} \right)^{2} + g_{1}{\sum\limits_{\mu = x,y}\left( {\psi^{\dagger}\tau_{\mu}\psi} \right)^{2}} + g_{2}\left( {\psi^{\dagger}\tau_{z}\psi} \right)^{2},
$$
with the coefficients 
$$
g_{0} = \frac{U_0 + g_{2}}{2},g_{1} = - \frac{u_{K} + g_{2}}{2},g_{2} = u_{N}.
$$
We note that, the $g_1$- and $g_2$-terms prefer the VBS and VP orders respectively, and 
then the VBS-VP transition boundary should be given by
\begin{equation}
g_1 = g_2,~g_1<0,~g_2<0 \quad\to\quad u_N = -\tfrac{1}{3}u_K,~u_N<0,~u_K>0,
\end{equation}
which is confirmed by our DMRG calculations in \Fig{fig:VP2VBS}.

For the case of FM to N\'eel transition, we should have an alternative form of the Hamiltonian. 
From the Fierz Identity \Eq{eq:Fierz} derived in \Sec{sec:III}, i.e., 
$$
\left( {\psi^{\dagger}O^{i}\psi} \right)^{2} = {\sum\limits_{j} \mp\frac{1}{4} {\left( {\psi^{\dagger}O^{j}\psi} \right)^{2} + \mu\psi^{\dagger}\psi}},
$$ 
we can derive
\begin{equation*}
(\psi^\dag \tau_x \psi)^2 + (\psi^\dag \tau_y \psi)^2 + (\psi^\dag \tau_z \psi)^2 + 
(\psi^\dag \sigma_x \psi)^2 + (\psi^\dag \sigma_y \psi)^2 + (\psi^\dag \sigma_z \psi)^2 + 
2(\psi^\dag\psi)^2 = 0,
\end{equation*}
which together with the relation derived in \Sec{sec:III}
$$
\left( {\psi^{\dagger}\tau_{z}\psi} \right)^{2} - {\sum\limits_{i = 1,2}\left( {\psi^{\dagger}\Gamma^{i}\psi} \right)^{2}} + \left( {\psi^{\dagger}\psi} \right)^{2} = - \left( {\psi^{\dagger}\tau_{z}\psi} \right)^{2} - {\sum\limits_{i = 3,4,5}\left( {\psi^{\dagger}\Gamma^{i}\psi} \right)^{2}} + 4\psi^{\dagger}\psi,
$$
gives the relation 
\begin{equation*}
3\sum\limits_{i=1,2}(\psi^\dag\Gamma^i\psi)^2 + 2\sum\limits_{i=1,2,3}(\psi^\dag \sigma_i \psi)^2 - \sum\limits_{i=3,4,5}(\psi^\dag\Gamma^i\psi)^2 + 3(\psi^\dag\psi)^2 =0
\end{equation*}
This leads to an alternative form of the Hamiltonian up to a chemical potential term, 
\begin{equation}
H = h_{0}\left( {\psi^{\dagger}\psi - 2} \right)^{2} + h_{1}{\sum\limits_{i=3,4,5}\left( {\psi^{\dagger}\Gamma^i\psi} \right)^{2}} + h_{2}\sum\limits_{i=1,2,3}\left( {\psi^{\dagger}\sigma_i\psi} \right)^{2},
\end{equation}
with the coefficients
\begin{equation}
h_0 = \frac{U_0-u_K}{2}, h_1 = -\frac{3u_N+u_K}{6}, h_2 = \frac{u_K}{3}.
\end{equation}
We note that, the $h_1$- and $h_2$-terms prefer the N\'eel and FM orders respectively, and 
then the FM-N\'eel transition boundary should be given by
\begin{equation}
h_1 = h_2,~h_1<0,~h_2<0 \quad\to\quad u_N = -u_K,~u_N>0,u_K<0,
\end{equation}
which is confirmed by our DMRG calculations in \Fig{fig:FM2Neel}.
}

\begin{figure}[h!]
	\includegraphics[width=.45\columnwidth]{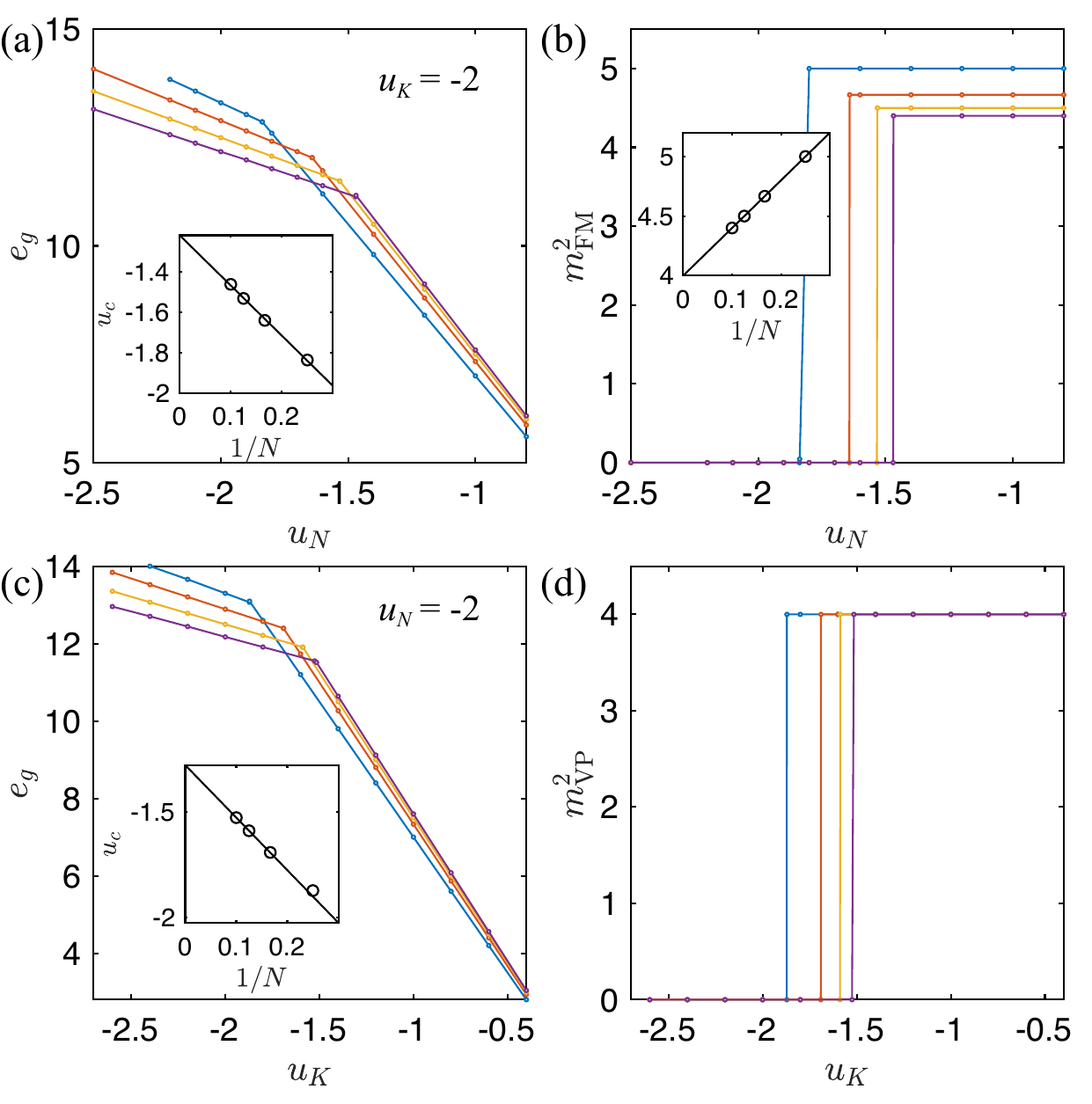}
	\caption{\textbf{More DMRG results for the FM and VP phases. }\\
Along a fixed $u_K=-2$ line, (a) the ground-state energies $e_g$ are shown versus $u_N$, and exhibit kinks behaviour for various system size. The kinks positions are extrapolated linearly versus $1/N$ to $u_c(N\to\infty)=-1.22$. 
(b) $m^2_\mathrm{FM}$ shows a sudden jump behaviour indicating the first order transitions. \\
Along a fixed $u_N=-2$ line, (c) the ground-state energies $e_g$ are shown versus $u_K$, and exhibit kinks behaviour for various system size. The kinks positions are extrapolated linearly versus $1/N$ to $u_c(N\to\infty)=-1.28$. 
(d) $m^2_\mathrm{VP}$ shows a sudden jump behaviour indicating the first order transitions. \\ 
	}
	\label{fig:FM_VP}
\end{figure}

\begin{figure}[h!]
	\includegraphics[width=0.65\columnwidth]{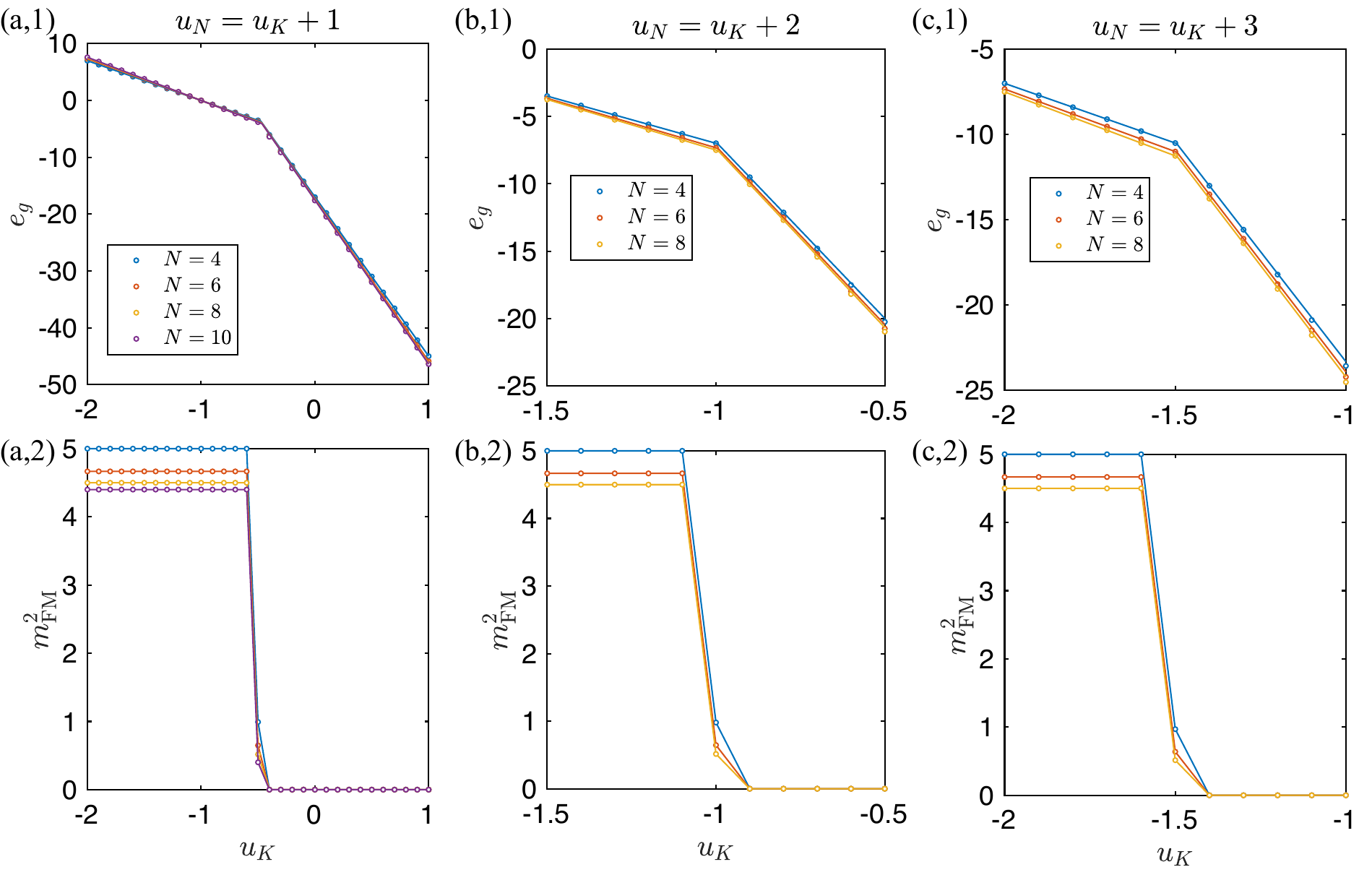}
	\caption{\textbf{More DMRG results for the FM-N\'eel transition. }\\
	For line $u_N = u_K +1$ in the phase diagram, 
	(a,1) the ground-state energies $e_g$ show kinks at $(u_K,u_N)=(-0.5,0.5)$ for various system sizes.
	(a,2) $m^2_\mathrm{FM}$ show finite value for $u_K<-0.5$ and suddenly drop to $0$ for $u_K>-0.5$. \\
	For line $u_N = u_K +2$ in the phase diagram, 
	(b,1) the ground-state energies $e_g$ show kinks at $(u_K,u_N)=(-1,1)$ for various system sizes.
	(b,2) $m^2_\mathrm{FM}$ show finite value for $u_K<-1$ and suddenly drop to $0$ for $u_K>-1$. \\
	For line $u_N = u_K +3$ in the phase diagram, 
	(c,1) the ground-state energies $e_g$ show kinks at $(u_K,u_N)=(-1.5,1.5)$ for various system sizes.
	(c,2) $m^2_\mathrm{FM}$ show finite value for $u_K<-1.5$ and suddenly drop to $0$ for $u_K>-1.5$.
	}
	\label{fig:FM2Neel}
\end{figure}

\begin{figure}[h!]
	\includegraphics[width=.8\columnwidth]{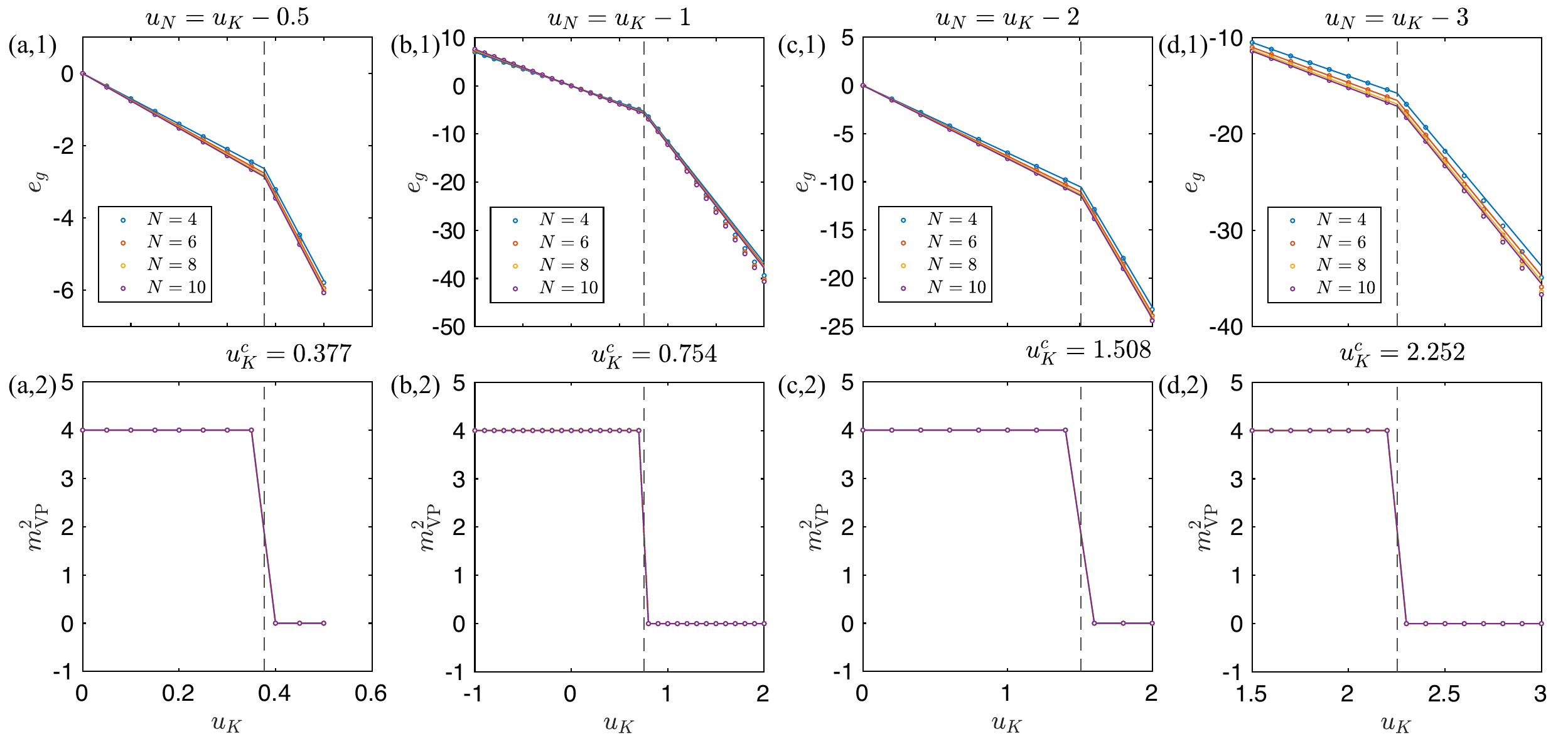}
	\caption{\textbf{More DMRG results for the VP-VBS transition. }\\
	For line $u_N = u_K -0.5$ in the phase diagram, 
	(a,1) the ground-state energies $e_g$ show kinks at $(u_K,u_N)=(0.377,-0.123)$ for various system sizes.
	(a,2) $m^2_\mathrm{VP}$ show finite value for $u_K<0.377$ and suddenly drop to $0$ for $u_K>0.377$. \\
	For line $u_N = u_K -1$ in the phase diagram, 
	(b,1) the ground-state energies $e_g$ show kinks at $(u_K,u_N)=(0.754,-0.246)$ for various system sizes.
	(b,2) $m^2_\mathrm{VP}$ show finite value for $u_K<0.754$ and suddenly drop to $0$ for $u_K>0.754$. \\
	For line $u_N = u_K -2$ in the phase diagram, 
	(c,1) the ground-state energies $e_g$ show kinks at $(u_K,u_N)=(1.508,-0.492)$ for various system sizes.
	(c,2) $m^2_\mathrm{VP}$ show finite value for $u_K<1.508$ and suddenly drop to $0$ for $u_K>1.508$. \\
	For line $u_N = u_K +3$ in the phase diagram, 
	(d,1) the ground-state energies $e_g$ show kinks at $(u_K,u_N)=(2.252,-0.748)$ for various system sizes.
	(d,2) $m^2_\mathrm{VP}$ show finite value for $u_K<2.252$ and suddenly drop to $0$ for $u_K>2.252$.\\
	Here the specific transition points are determined by linear extrapolation (indicated by the solid lines) 
	of data near the transitions.
	}
	\label{fig:VP2VBS}
\end{figure}

\end{widetext}
\end{document}